# Investigating the Behavior and Spatiotemporal Variations of Green-line Emission in the Solar Corona

Jacob Oloketuyi[1,2], Yu Liu[1], Linhua Deng[3], Abouazza Elmhamdi[4], Fengrong Zhu[1], Ayodeji Ibitoye[5,6,7], Opeyemi Omole[2], Feiyang Sha[1], and Qiang Liu[8]

[1] School of Physical Science and Technology, Southwest Jiaotong University, 999, Xi'an Road, Pidu District, Chengdu 611756, Sichuan, People's Republic of China; lyu@swjtu.edu.cn, jacob.oloketuyi@swjtu.edu.cn
[2] Department of Physics, Bamidele Olumilua University of Education, Science and Technology, Ikere-Ekiti City, P.O. Box 250, Ekiti State, Nigeria
[3] School of Mathematics and Computer Science, Yunnan Minzu University, Kunming 650504, Yunnan, People's Republic of China
[4] Department of Physics and Astronomy, King Saud University, Riyadh, 11451, Saudi Arabia
[5] National Astronomical Observatories, Chinese Academy of Sciences, 20A Datun Road, Chaoyang District, Beijing 100101, People's Republic of China
[6] Centre for Space Research, North-West University, Potchefstroom 2520, South Africa
[7] Department of Physics and Electronics, Adekunle Ajasin University, P. M. B. 001, Akungba-Akoko, Ondo State, Nigeria
[8] School of Physics and Astronomy, China West Normal University, Nanchong, Sichuan, People's Republic of China

## Abstract

Understanding coronal structure and dynamics can be facilitated by analyzing green-line emission, which enables the investigation of diverse coronal structures such as coronal loops, streamers, coronal holes, and various eruptions in the solar atmosphere. In this study, we investigated the spatiotemporal behaviors of green-line emissions in both low and high latitudes across nine solar cycles, ranging from Solar Cycle 17 to the current Solar Cycle 25, using the modified homogeneous data set. We employed methodologies such as cross correlation, power spectral density, and wavelet transform techniques for this analysis. We found distinct behaviors in green-line energy across various latitudinal distributions in the solar atmosphere. The trends observed at higher latitudes differ from those at lower latitudes. The emission behaviors show a close association with other solar phenomena like solar flares, sunspots, and coronal mass ejections throughout the solar cycles. The observed variations exhibit harmonic periods. The emission activity is significantly higher in the low latitudes, accounting for over 70% of the emissions, while the higher latitudes contribute less than 30%. The emissions exhibit asymmetric behavior between the northern and southern hemispheres, leading to a 44 yr cycle of solar hemispheric dominance shifts. Various factors, such as Alfvén waves, solar magnetic fields, sunspots, differential rotation, and reconnection events, influence the observed differences in behavior between lower and higher latitudes, suggesting the existence of potential underlying phenomena contributing to deviations in properties, intensity, temporal dynamics, and spatiotemporal lifetime.

## 1. Introduction

The Sun, the nearest star to us, serves as a significant source of energy that drives numerous phenomena within the solar system and beyond. The solar corona, the ever-changing section of the Sun's outermost layer, is pivotal for these processes. Studying the green line (530.3 nm emission of Fe XIV) is vital, as it provides essential insights into the thermal structure, density, and magnetic fields of the corona. This understanding is crucial for grasping solar activities and dynamics in the solar atmosphere (S. R. Habbal et al. 2011). The characteristics of the green line's intensity and emission are fundamental in investigating temperature distributions, calculating electron densities, and examining coronal magnetic fields (O. G. Badalyan & V. N. Obridko 2006; S. R. Habbal et al. 2009; B. Boe et al. 2020). This emission line is significant for observing solar flares and coronal mass ejections (CMEs), aiding in space-weather prediction and addressing the coronal heating issue (J. Takalo 2022). Additionally, the extensive record of green-line observations provides strong evidence for long-term solar cycle research, spanning up to 85 yr covering nine solar cycles in this study, thereby validating the solar observation instruments and making it an indispensable tool in solar physics.

This part of the solar atmosphere emits light in narrow spectral emission lines when solar magnetic energy is released in flares and CMEs. The green "530.3 nm" line is emitted from a million-degree Fe XIV ion layer. Prior observations suggest changes in the structure and brightness of the green-line corona following some flares and CMEs, but the nature of these changes and their physical mechanisms are still poorly understood (A. Warmuth et al. 2004; B. N. Dwivedi & A. K. Srivastava 2006; I. Suzuki et al. 2006; K. Shibata & T. Magara 2011; E. Antonucci et al. 2020).

To gain a comprehensive understanding of the behavior and spatiotemporal relationships linking the physical mechanisms behind eruptions and heating of the corona, this study will investigate green-line emission activities throughout the entire 360° of the solar atmosphere, in conjunction with CME events at multiple emission angles. Previously, D. Moses et al. (1997) observed a CME initiating with Fe XII. Additionally, phenomena such as filament eruption, coronal vacuum formation, the beginning of a brilliant outward-moving shell, and a more distant wave have been documented (D. Moses et al. 1997;

S. P. Plunkett et al. 1997). The position of an associated phenomenon must be nearly identical to that of its coronal transient (R. H. Munro et al. 1979). By studying the interrelationships between solar activities such as CMEs, coronal holes, and green-line corona intensity, we can better understand solar magnetic activity through coronal seismology. As I. Ermolli et al. (2014) mentioned, utilizing various solar parameters in the study of solar dynamics, intensity, and fluctuations offers a better understanding of solar magnetic configurations and eruptions. Furthermore, B. Schmieder et al. (2014) and B. Schmieder & G. Aulanier (2018) observed that magnetic flux emergence alone is not sufficient for eruptions, while A. O. Benz & M. Gudel (2010) provided additional insights by stating that a significant portion of the energy converts into nonthermal particles, which are then disseminated to denser gas, heating it to temperatures that emit soft X-rays.

Previous studies have provided significant insights at various levels of investigation. This work builds on several earlier scientific initiatives to understand different aspects of the solar atmosphere, especially the corona, including, but not limited to, T. Gold & F. Hoyle (1960), M. E. Machado et al. (1978), S. F. Martin (1980), K. Kusano et al. (2004), H. Song et al. (2009), G. Porfir'eva et al. (2012), J. T. Karpen et al. (2012), H. Q. Song et al. (2013), I. Ermolli et al. (2014), J. Chen (2017), P. Bhowmik & A. R. Yeates (2021), and N. Xiang et al. (2023). A recent study conducted by A. S. H. To et al. (2023) shows that variations in coronal abundances are significantly impacted by the intensity of the magnetic field in active areas. This is thought to be the main reason for the nonlinearity when solar activity is at its highest. Y.-M. Wang & S. Hawley et al. (1997) discovered that the poleward concentration of the large-scale photospheric field occurs mostly at solar minimum and that there are increases in green-line emission at high latitudes. The study also found that magnetic flux reconnections from complicated solar active areas with high-latitude unipolar fields are closely associated with the brightest increases.

The correlation between sunspot regions and green-line strength is contingent on the solar cycle phase, since varying magnetic field scales influence the development of distinct physical conditions inside the inner solar corona, which in turn triggers the emergence of green-line occurrences (Y.-M. Wang et al. 1996; K. Wilhelm et al. 1998; N. G. Bludova & O. G. Badalyan 2006; J.-E. Hwangbo et al. 2014). However, O. Badalyan (2013) discovered evidence that the intricate nature and influence of various fields have an impact on coronal heating processes and structure formation, which lead to the emergence of Fe XIV ions and the emission of the green coronal line, which are dependent on specific amounts of coronal magnetic field structures.

Analyzing the latitudinal distribution of green-line corona emissions, N. G. Bludova et al. (2014) found the relationship between the green coronal line intensity and sunspot areas and magnetic fields using data between 1977 and 2001. The authors showed that the correlation between the green-line intensity and sunspot areas decreases at higher-latitude zones. The study also revealed that all the correlation coefficients obtained are naturally associated with the 11 yr cycle of solar activity. O. G. Badalyan & V. N. Obridko (2006), using the magnetic field's strength and its tangential and radial components, respectively, obtained correlation coefficients from pair-to-pair differences in latitude and time showing different levels of relationships with magnetic fields. The findings of N. Xiang et al. (2023) demonstrate the impact of various magnetic structures in the rotation of the coronal atmosphere; the changes in the rotation of the plasma environment in the corona are influenced mainly by small-scale magnetic components.

This work aims to comprehensively evaluate and explain the distribution of solar magnetic energy across different latitudes by analyzing green-line corona emission associated with solar energy. The decision to utilize green-line emission instead of H$\alpha$ stems from its exceptional suitability for investigating the brightness and temperature sensitivity of the corona, enabling the evaluation of the various structures within the solar atmosphere (S. R. Habbal et al. 2011; H. T. Li et al. 2023). This project aims to thoroughly investigate solar activity in the corona and the complex relationships between solar flares and CMEs, utilizing green-line corona emissions. The minor objectives involve analyzing the spatiotemporal patterns over nine solar cycles. This will be done by studying the differences in green-line emission between the northern and southern hemispheres, as well as investigating temporal variations and connections in solar corona activity. The ultimate aim is to enhance our understanding of coronal heating and ionization processes by gaining valuable insights. Our study is presented in four parts: introduction, data and methods, results, and conclusions.

## 2. Data and Methods

### 2.1. Data

The green coronal intensity is the spectral line, which is the emission of excited and ionized iron atoms that provide the radiant energy released by the entire visible corona. The modified homogeneous data set (MHDS) was used in carrying out this study and was obtained as a result of modifications to the homogeneous data set (HDS), which includes a combination of satellite- and ground-based data. The intensity scale for the 530.3 nm (Fe XIV) green coronal line observed at ground-based coronal stations up to 2008 was standardized under the name HDS. Starting from 1996, the MHDS data were utilized in place of the HDS data to enhance the density of observations and eliminate the necessity for interpolation (M. Rybanský et al. 2005). The MHDS of coronal intensities integrates data from both ground-based and space-borne instruments to provide a comprehensive record of solar activity. Ground-based observations of the green coronal emission line (530.3 nm) were obtained using various coronagraphs worldwide, forming the basis of the HDS. Space-borne data were primarily gathered from the Extreme-ultraviolet Imaging Telescope (EIT) on the Solar and Heliospheric Observatory (SOHO) satellite, which captured high-resolution images of the Sun in the 28.4 nm Fe XV line. Additionally, the Charge Element and Isotope Analysis System (CELIAS) on SOHO provided measurements in the 26–34 nm region, which were used to develop the modified coronal index (MCI; M. Rybanský et al. 2005; I. Dorotovič et al. 2014).

To ensure consistency and accuracy, a calibration procedure was employed to correlate the coronal index (CI) from HDS with mean intensity values from daily observations, achieving a significant correlation coefficient. Relative intensities from EIT images were adjusted to match the intensity scale of the Fe XIV line, and interpolation methods were used for periods without direct measurements to maintain data continuity. The EIT images have a spatial resolution better than $1'$ and are taken using an array of 1024 × 1024 pixels (J. Zhu et al. 2021). This comprehensive approach permits a detailed analysis of solar activity over time. It leverages historical data and modern observations to create a valuable tool for studying the evolution of the solar corona in spatial and temporal dimensions.

To ensure consistent and accurate measurements of coronal intensities, data from various instruments and observational conditions were processed using a number of procedures, including the weighted least-squares technique, which allowed the creation of the HDS, which provides a uniform data set despite variations in observational circumstances and devices (M. Rybanskỳ et al. 2005). From the HDS, the CI was constructed, which represents the total irradiance from the entire corona in the 530.3 nm line.

Additionally, the CI was extended to create the MCI by incorporating measurements from the CELIAS on board SOHO (I. Dorotovič et al. 2014). The correlation between the CI and mean intensity was then determined, using the average of 72 daily HDS values, to ensure compatibility. This correlation was found to be significant at $r = 0.8986$ based on selected observation days from the Lomnický Štít observatory (I. Dorotovič et al. 2014). The correlation functions derived from this analysis served as the basis for calibrating the MHDS (M. Rybansky 1975; M. Rybansky & V. Rusin 1985, 1992; I. Dorotovič et al. 2014, and M. Rybanskỳ et al. 2005).

The reconstructed MHDS data are measured daily into 5° measuring position angle intervals starting from the northern pole and moving toward the solar eastern region, such that 0°, 90°, 180°, and 270° are measured for the solar north, east, south, and west, respectively. The measurement is taken across the solar surface. We then reclassified into northern and southern hemispheres, which are further reconstructed into lower and higher latitudes. The reconstruction is as follows: the northern region includes measurements from 0° to 90° and from 270° to 360°. The south includes 90°–270°. The high-latitude measurements include 0°–40°, 320°–360° in the north, and then 140°–220° in the south. The low latitudes include measurements from 50° to 130° and from 230° to 310°. Measurements taken at borderlines were excluded, with angles such as 45°, 135°, 225°, and 315°. The equator is represented by the straight line that runs from 90° to 270°. The data span from 1939 January to 2024 March, covering nine solar cycles, from Solar Cycle 17 to the current Solar Cycle 25. Detailed information about the measurement, instruments, and observatories is provided by M. Rybansky (1975), I. Dorotovič et al. (2014), and M. Rybanskỳ et al. (2005) and can be downloaded from the website.[9]

## 2.2. Methods

### 2.2.1. The North–South Asymmetry

The N–S asymmetry is introduced to measure the uneven evolution of green-line emissions in the solar atmosphere. The N–S asymmetry of the corona of the Sun pertains to the observed disparities between the atmospheric regions of the Sun's northern and southern hemispheres and can be presented as

$$C = \frac{C_N - C_S}{C_N + C_S}, \tag{1}$$

where $C_N$ and $C_S$ represent the observed green-line emissions obtained from the northern and southern hemispheres, respectively (G. Vizoso & J. Ballester 1990; T. Ataç & A. Özgüç 1996; K. Li et al. 2002).

### 2.2.2. The Cross Correlation

Cross correlation was employed in this study to assess the magnitude and direction of the association between the green corona evolutions in the southern and northern regions of the Sun. The correlation coefficient value is dimensionless and ranges from −1.0 to +1.0 ratio of variances, with wide industrial applications. The cross correlation between the corona emission in the northern and southern hemispheres can be defined as

$$CC(\Delta) = \frac{\sum_{i=1}^{n}[C_N(i) - N]\,[C_S(i + \Delta) - S]}{(n - 1)\delta_N\,\delta_S}. \tag{2}$$

The variables $N$ and $S$ represent the average values for the northern and southern hemispheres, respectively, while the variables $\delta_N$ and $\delta_S$ correspond to their respective standard deviations (M. Hagino et al. 2004).

### 2.2.3. The Power Spectral Density

Power spectral density (PSD) describes the distribution of power over frequency computation with the Fourier transform of a time-domain data stream. The theorem also holds in discrete-time cases, as described by R. N. Youngworth et al. (2005), P. Andren (2006), and R. H. Shumway et al. (2000). The temporal fluctuations, periodicities, and distinct peak features exhibited by coronal emissions across the solar cycles can be analyzed using PSD, which can provide valuable insights. If the green-line emission data set has $N$ continuously recorded values $z(t)$, $z(t)$ represents the emission of the green line as a function of length $t$, and a Fourier transform with finite length may be expressed as

$$Z(t) = \int_0^L dt\; z(t)\exp(-ikt), \tag{3}$$

where $t$, $\Delta t$, and $L = N\Delta t$ are given by the wavenumber, space interval, and total length, respectively. Details of this process and procedure are provided in J. M. Elson & J. M. Bennett (1995) and R. H. Shumway et al. (2000).

### 2.2.4. The Continuous Wavelet Transform

The continuous wavelet transform (CWT) is a quantitative depiction, which is a scalogram or wavelet power spectrum (WPS), that illustrates energy distribution in a signal across temporal and frequency dimensions. This method was utilized to identify temporary occurrences, specific characteristics, and variations in the frequency of green-line emission that may not be apparent in conventional Fourier analysis (C. Torrence & G. P. Compo 1998; J. Polygiannakis et al. 2003; A. Grinsted et al. 2004). The CWT signal $x(t)$ as presented by C. K. Chui (1992) is

$$X(b, a) = |a|^{-1/2}\int_{-\infty}^{+\infty} x(t)\bar{\varphi}\left(\frac{t - b}{a}\right)dt, \tag{4}$$

[9] https://www.kozmos-online.sk/slnko/modifikovany-homogenny-rad-modified-homogeneous-data set/

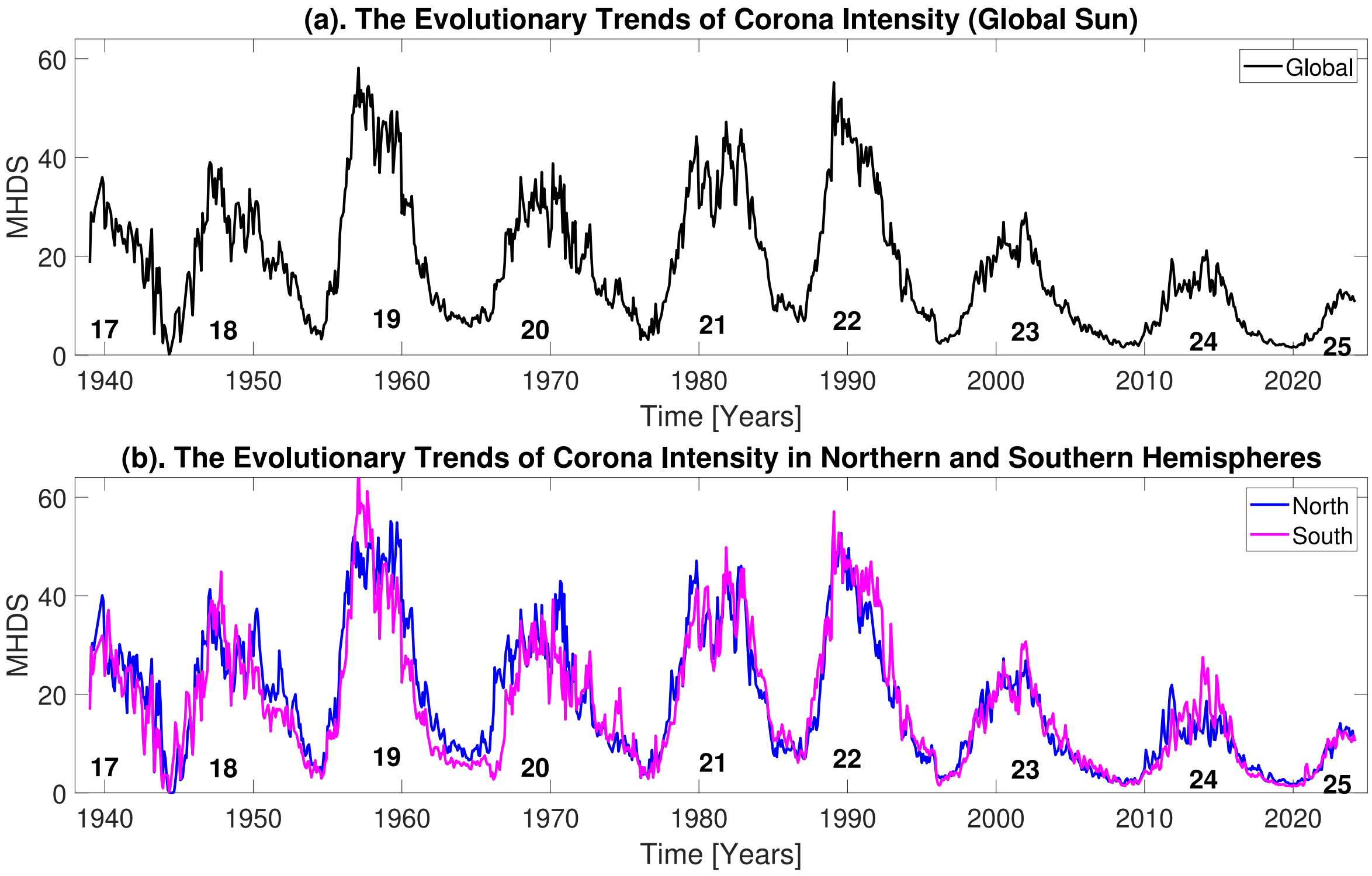

**Figure 1.** The evolutionary trends of green-line coronal emissions from 1939 to 2024. Panel (a) shows the global trend in black, while panel (b) shows the trends for the northern (blue) and southern (magenta) hemispheres.

where $\varphi$ is the Morlet wavelet function (H. Schmitz-Hübsch & H. Schuh 2003) and is defined as

$$\varphi(t) = \frac{\exp(ipt)}{\sqrt{2\pi}}\left[Q(t, \sigma)(1 - \sqrt{2})\exp\left(\frac{-p^2\sigma^2}{4}\right)\right], \quad (5)$$

where we have imposed the transformation

$$Q(t, \sigma) = \exp\left(\frac{-t^2}{2\sigma^2}\right). \quad (6)$$

Here $\sigma$ is the Morlet decay parameter, $p$ (typically >5) is the frequency parameter, $a$ is the dilation parameter ($a \neq 0$), and $b$ is the translation parameter. For $p = 2\pi$ the dilated Morlet wavelet oscillation period equals $a$ (W. Popiński & W. Kosek 1994).

## 3. Results and Discussions

### 3.1. The Spatiotemporal Variations

The evolutionary trends of green-line emissions are presented in Figure 1. Panel (a) shows the global trend for the overall period of this investigation. The data present the monthly average of green-line coronal emission between 1939 January and 2024 March. It is observed that Solar Cycle 19 has the highest values, followed by Solar Cycle 22, while Solar Cycles 23 and 24, as well as Solar Cycle 25, have lower emissions of the green line. As recently observed by J. Oloketuyi et al. (2024), the green-line emissions show a continuous waning. This is shown in Solar Cycle 25, which is already moving toward the descending phase of the cycle. The violation of the regular alternation of cycle heights, known as the Gnevyshev–Ohl rule (GOR), is also observed in Figure 1. This rule, which states that an even cycle is lower than the subsequent odd one, holds for pairs 18–19 and 20–21 but is violated in pairs 22–23 and 24–25. The GOR is fulfilled in the sunspot data, as recently published by Y. A. Nagovitsyn et al. (2024). This violation in the green-line coronal data is particularly interesting.

Panel (b) presents the trends for evolutions across the northern and southern hemispheres. It is observed that the evolutions in both hemispheres almost followed the same pattern. In Solar Cycle 19, the northern emissions exhibit double-peak features (Gnevyshev gap), while the southern emissions are higher during the peak period and lower in the descending phase, continuing toward the rising phase of Solar Cycle 20.

### 3.2. The Latitudinal Distributions of Green-line Coronal Emissions

The evolutionary trends and latitudinal distributions of green-line coronal emissions are illustrated in Figures 2–5. This was introduced to reveal distinct behaviors exhibited at lower and higher latitudes in both the northern and southern solar regions. Figure 2(a) depicts the global trends across various latitudes from the northern and southern hemispheres of the Sun. Emission rates are higher at lower latitudes compared to higher latitudes. Additionally, it is noted that the periodicity at lower latitudes does not necessarily match the patterns seen at higher latitudes throughout the solar cycles. During the rising phase of Solar Cycles 18, 19, and 22, an early increase in green-line emissions is observed at lower latitudes. As previously noted, the green-line coronal activity is reduced in Solar Cycles 23, 24, and 25, at both lower and higher latitudes. The latitudinal analysis for the northern hemisphere is detailed in Figure 2(b), showcasing the features and characteristics at low and high latitudes. Solar Cycles 19, 21, and 22 demonstrate higher green-line coronal activity, whereas lower activities are found in Solar Cycles 23, 24, and 25. Early increases in green-line emissions are again observed at low latitudes, especially during Solar Cycles 18, 19, and 22. In Solar Cycles 17, 20, and 24, activity is notably low at high latitudes, while the concluding phases of Solar Cycles 18 and 22 are lower than their rising phases. This trend is similarly observed in the southern hemisphere, as shown in Figure 2(c).

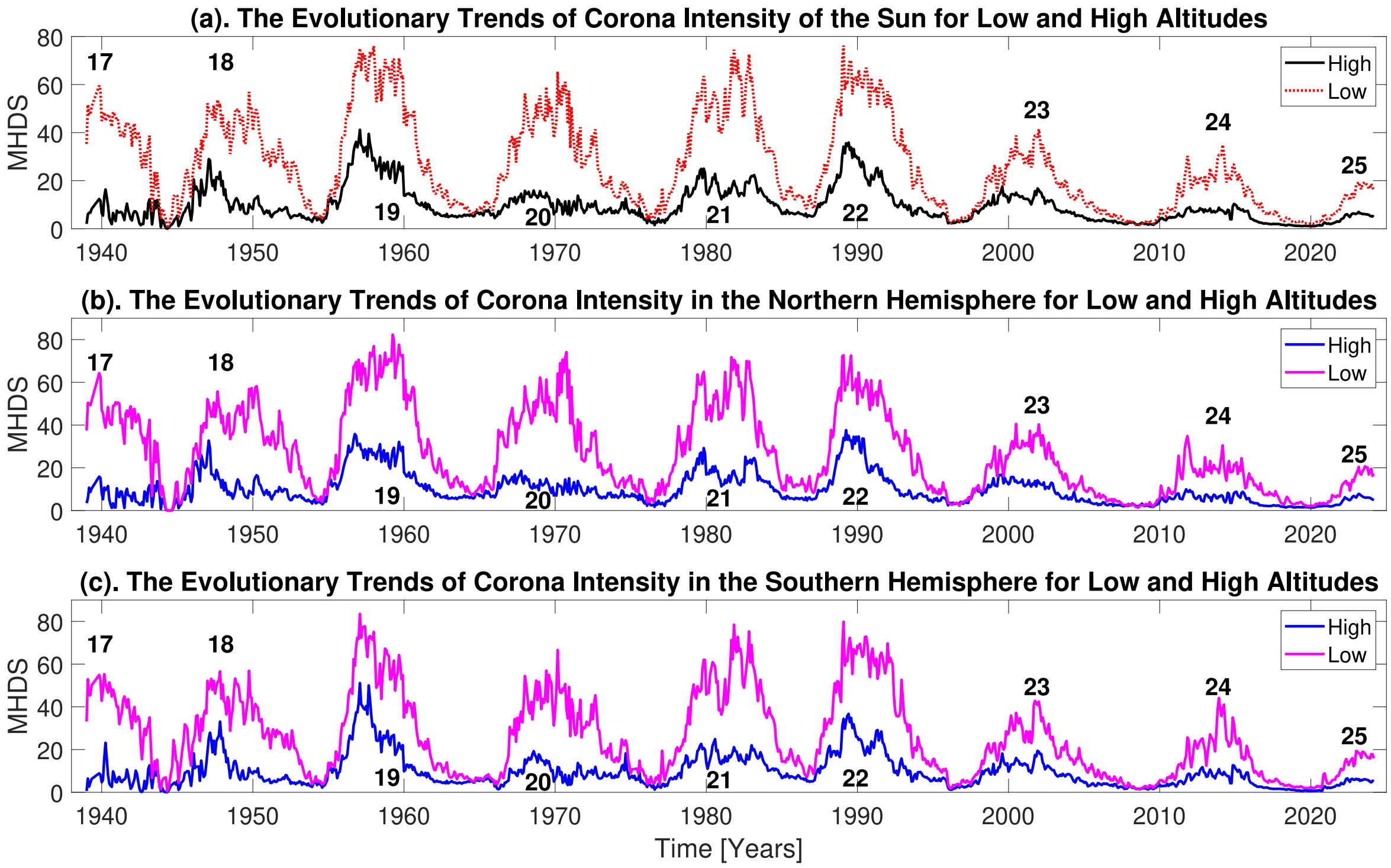

**Figure 2.** The evolutionary trends obtained from low and high latitudes of the Sun. Panel (a) shows the global trends for high (black) and low (red) latitudes. Panels (b) and (c) show trends obtained from the northern and southern hemispheres, respectively, where blue and magenta colors corresponds to high and low latitudes, respectively.

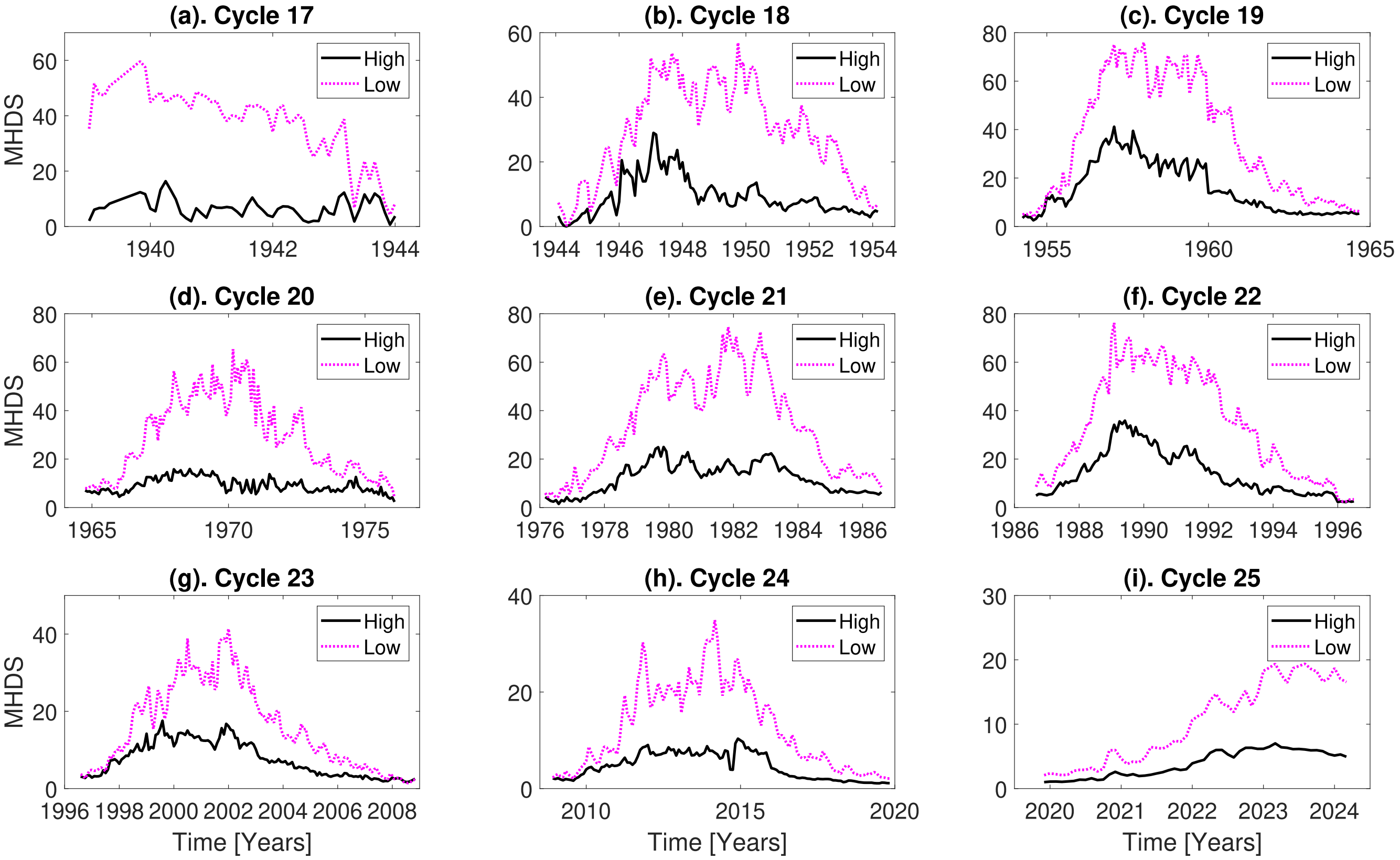

**Figure 3.** The evolutionary trends obtained for high (black) and low latitudes (pink) for solar cycles.

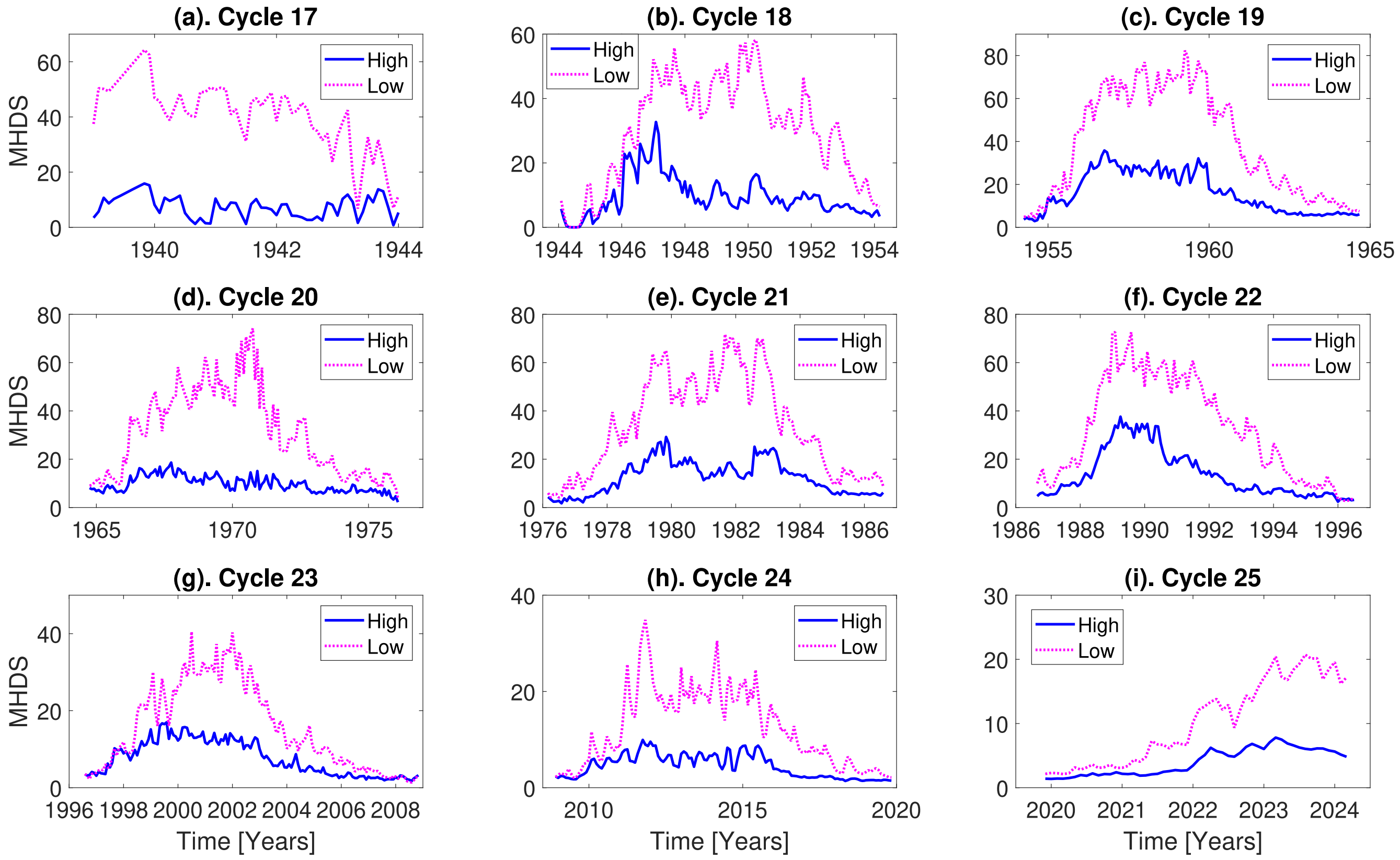

**Figure 4.** The evolutionary trends in the northern hemisphere obtained for high (blue) and low (pink) latitudes.

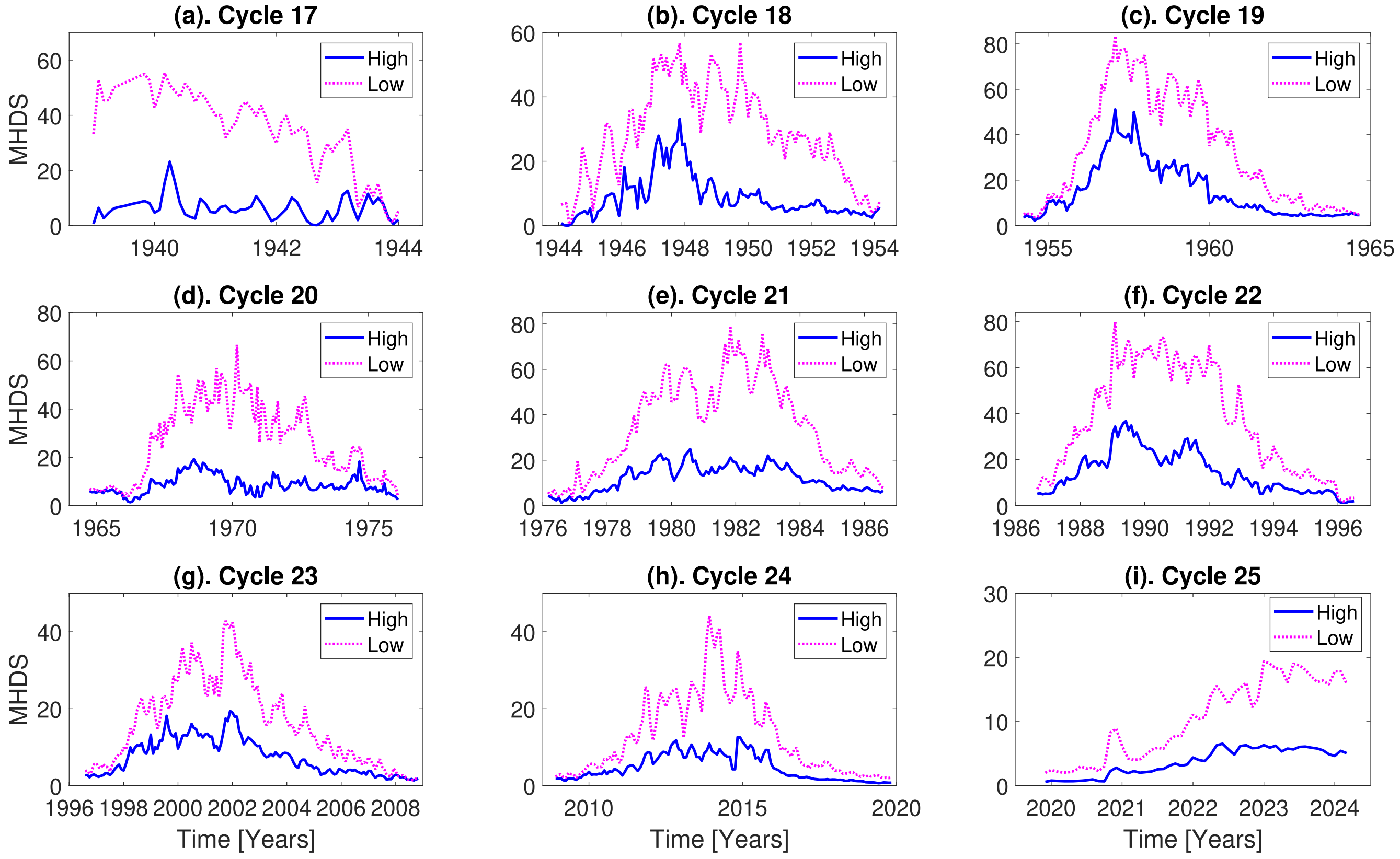

**Figure 5.** The evolutionary trends in the southern hemisphere obtained for high (blue) and low (pink) latitudes.

**Table 1**
The North–South Asymmetry, Showing the Means and the Dominance Region

| Global Scale | North Mean | South Mean | N–S Asymmetry | Dominance |
|---|---|---|---|---|
| Total (Global) | 18.61 | 17.85 | 0.76 | North |
| Global High | 9.90 | 9.50 | 0.40 | North |
| Global Low | 27.32 | 26.20 | 1.13 | North |

| | Total Cycle | | | | Rising Phase | | | | Declining Phase | | | |
|---|---|---|---|---|---|---|---|---|---|---|---|---|
| Cycle | North | South | Asym. | Dom. | North | South | Asym. | Dom. | North | South | Asym. | Dom. |
| 17 | 24.20 | 21.19 | 3.01 | N | 32.37 | 27.81 | 4.56 | N | 22.40 | 19.74 | 2.67 | N |
| 18 | 19.84 | 18.71 | 1.13 | N | 17.67 | 19.19 | −1.52 | S | 21.15 | 18.41 | 2.73 | N |
| 19 | 27.55 | 23.86 | 3.69 | N | 31.94 | 31.34 | 0.60 | N | 24.56 | 18.77 | 5.79 | N |
| 20 | 20.45 | 17.57 | 2.88 | N | 23.83 | 18.46 | 5.37 | N | 17.30 | 16.74 | 0.56 | N |
| 21 | 23.19 | 23.34 | −0.14 | S | 19.60 | 17.15 | 2.45 | N | 25.33 | 27.02 | −1.68 | S |
| 22 | 23.15 | 25.69 | −2.54 | S | 28.45 | 29.89 | −1.43 | S | 20.26 | 23.41 | −3.15 | S |
| 23 | 11.55 | 11.82 | −0.27 | S | 14.85 | 13.85 | 1.00 | N | 9.30 | 10.44 | −1.14 | S |
| 24 | 8.44 | 8.78 | −0.34 | S | 9.85 | 9.59 | 0.27 | N | 7.23 | 8.09 | −0.86 | S |
| 25 | 7.15 | 7.06 | 0.09 | N | 5.62 | 5.76 | −0.15 | S | 12.25 | 11.39 | 0.86 | N |

It is also noted that emission activity at high latitudes follows a similar rising pattern.

The trends for solar cycles are presented in Figure 3. The black and pink colors represent the high and low latitudes, respectively. As seen in the figure, the green-line emission activity at the high and low latitudes is distinct. The emission at high latitudes displays peaks that are dissimilar to those at low latitudes. For instance, in Solar Cycle 17, the emissions exhibit a similar pattern at high and low latitudes between 1939 and 1940, after which they follow a different trend. In Solar Cycles 17, 18, and 20 the green emissions at high latitude are weaker from the peak time to the descending phase of the cycles. Figure 4 shows the evolutionary trends in the northern hemisphere. The blue color represents the emissions at high latitudes, while the pink color corresponds to green-line emissions at low latitudes. Similar to observations in Figure 3, the emissions at high latitudes exhibit weaker activity in Solar Cycles 17, 18, and 20. It is also noticeable that, at high latitudes, the green-line emissions exhibit higher activity at the ascending phases of Solar Cycles 18, 19, 22, and 23.

The activity of green-line emission in the southern region is presented in Figure 5. As was observed in the north, the emission's activity shows weaker activity at the descending phase of Solar Cycles 17, 18, 20, 22, and 24. Generally, the emission activity at low latitudes is observed with more peaks and more strength than the activity at high latitudes, which tend to exhibit more irregular patterns.

### *3.3. The North–South Asymmetry*

The north–south asymmetry is presented in Table 1. This was introduced to measure the uneven evolution of green-line emissions across the northern and southern parts of the solar atmosphere. The table is divided into upper and lower parts. The upper part presents the asymmetry for the global scale, while the lower part presents the total cycle, rising phase, and declining phase. The table presents the mean average values for the north, south, asymmetry, and dominance regions. The analysis shows that the northern region dominates on the global scale with tiny margins at 0.76, 0.40, and 1.13 for the global, high, and low latitudes, respectively. The northern domination is shown from Solar Cycle 17 straight to Solar Cycle 20 before taking a southward turn to southern dominance for another four straight solar cycles from 21 to 24. The dominance in Solar Cycle 19 records the highest lead with 3.69, followed by Solar Cycle 17 with 3.01. The lowest lead is recorded in the current Solar Cycle 25, which is still an ongoing process and could be flipped to either side. It is observed that Solar Cycle 19 has the highest values of emission for both the northern and southern hemispheres at 27.55 and 23.86, respectively.

The analysis of the hemispheric distributions of green-line emissions is further scrutinized to show the pattern of dominance in Figure 6. As seen in Figure 6(a), the paradigm of N–S dominant shifting from north to south at intervals of four solar cycles demonstrates a harmonic 22 yr solar magnetic cycle. In Figure 6(b), the total average means for the northern and southern hemispheres are analyzed. These average mean values are presented in the lower part of Table 1. The northern mean values (red plot) led from Solar Cycle 17 to 20, before the south (black plot) took the lead from Solar Cycle 21 to 24, thereby showing the uneven behavior between the northern and southern hemispheres. This results in a 44 yr shift in solar hemispheric dominance. It is worth noting that 85 yr of series data is only partially adequate for a reliable conclusion.

To better understand the emission behavior during different solar cycle phases (rising and declining), we divided each cycle into these two phases. The analysis is presented in Figures 6(c) and (d). Figure 6(c) presents the asymmetries for both hemispheres during the rising and declining phases. In the rising phase, the northern hemisphere leads the southern hemisphere, except in Solar Cycles 18, 22, and 25. The declining phase reflects the observations for the entire cycle, where the northern hemisphere leads the southern hemisphere in Cycles 17−20 and Cycle 25. Investigating the hemisphere-specific emission activities throughout the different phases of the solar cycles, the average values were analyzed and are shown in Table 1 and Figure 6(d). The average emission in the rising phase of the northern hemisphere (NR) is higher in six cycles (17, 19, 20, 21, 23, and 24), and during the declining phase (ND) it is higher in five cycles (17, 18, 19, 20, and 25). In contrast, the average emissions in the southern hemisphere's rising and declining phases (SR and SD) are generally lower. The southern hemisphere leads only in three cycles (18, 22, and 25) during the rising phase and in four cycles (21, 22, 23, and 24) during the declining phase.

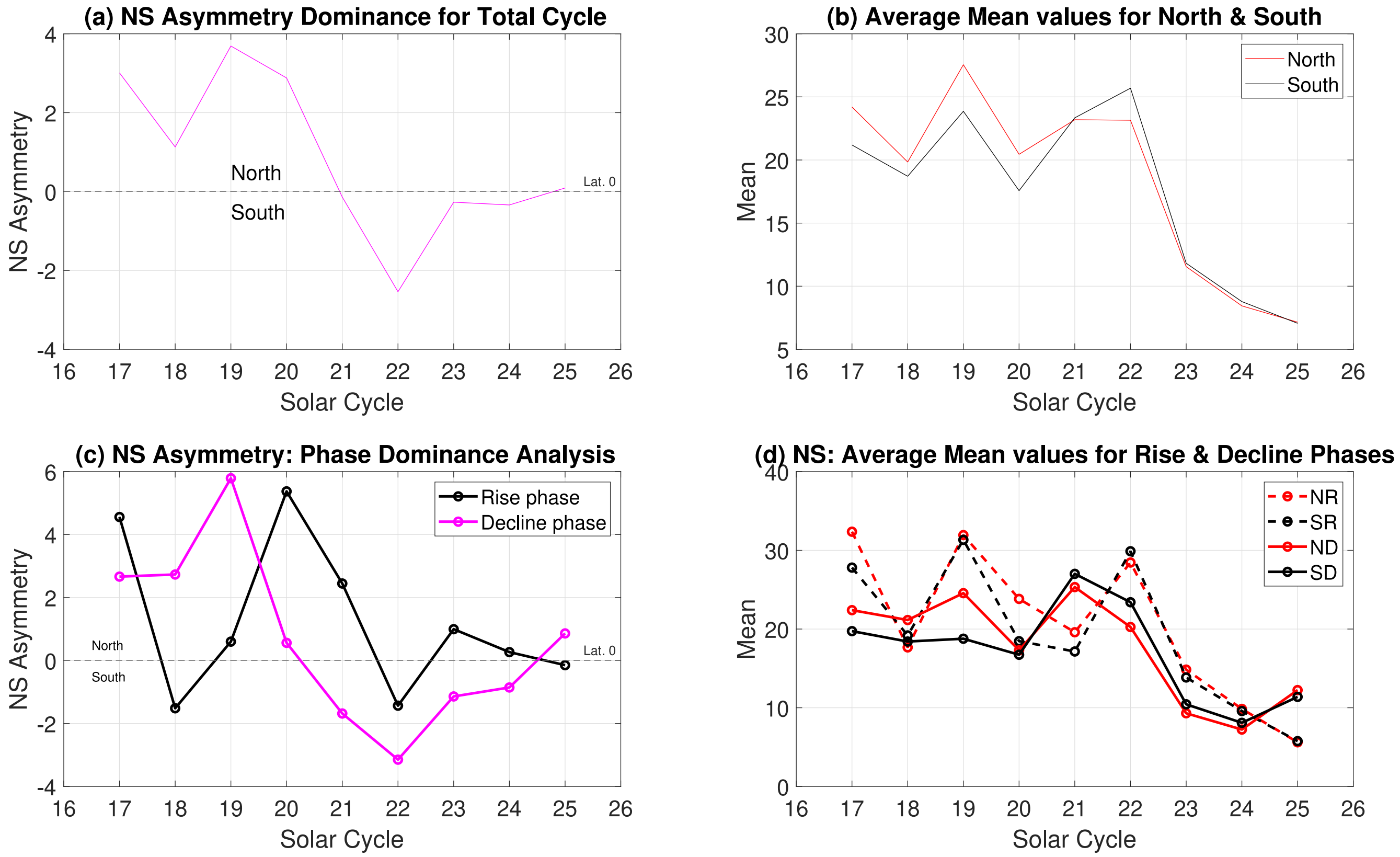

**Figure 6.** The north–south asymmetry. (a) The dominance trend showing the leading region. (b) Average mean values for north and south. (c) North–south asymmetry showing the leading phase across the northern and southern hemispheres. (d) Average mean values for rising and declining phases across north and south, where NR and SR represent the rising phase for the northern and southern hemispheres, while ND and SD represent the declining phase for the north and south, respectively.

Interestingly, the lead in Solar Cycle 18 by the northern hemisphere was largely due to emissions during the declining phase of that cycle. Similarly, the lead observed for the southern hemisphere in Solar Cycles 23 and 24 was primarily driven by emissions during their respective declining phases. A similar pattern is currently observed, with the northern hemisphere leading in Solar Cycle 25. In summary, green-line emissions during the declining phases are the primary drivers of the total leads observed, significantly influencing the outcomes in Solar Cycles 18, 19, 21, 22, 23, 24, and 25.

The asymmetric behavior also reflects variations in the solar cycle phases. The harmonic cycle analysis of the corona intensity data reveals the underlying periodicity and interactions within the solar cycles. In studying the solar flare classes between Solar Cycles 21 and 25, H. T. Li et al. (2023) observed an overall southern dominance of flares for all classes (B, C, M, X). This is in agreement with our observation. In the investigation of the solar transition region, Q.-R. Wu et al. (2023) observed dominance of the southern hemisphere between 2011 and 2022, indicating that the rotational speed of the southern part is somewhat higher than that of the northern hemisphere, with a difference of around 0.44%. Several studies have documented comparable findings of N–S dominance behavior (D. H. Hathaway & R. M. Wilson 1990; H. Xu et al. 2021; M. Wan & P.-x. Gao 2022). J. Sharma et al. (2021), studying the solar time series using the sunspot, attributed N–S asymmetry to inherent nonlinear dynamics, which are distinguished by the intricate nature of nonlinear phenomena.

### 3.4. The Cross Correlation

The technique was implemented to illustrate the degree of connection between the green-line emission activity in the northern and southern regions of the Sun. This analysis is displayed in Figures 7 and 8. Detailed information can be found in Table 2. The correlation coefficients measured are notably high, indicating strong relationships. Figure 7 presents the analysis for the global Sun recording a 0.93 correlation coefficient without recording a lag or lead, demonstrating a high level of phase synchronizations between emissions from both regions. The correlation coefficient obtained for high latitudes shows 0.06 less than the global value, while the analysis at low latitudes returns 0.01 less than the global Sun, all without lag or lead recorded. However, the cross-correlation figure also illustrates the asymmetry in the correlation curves. In the plots, the right wing (positive time lags) differs from the left wing (negative time lags), which indicates a distinct temporal relationship between the emissions from the northern and southern hemispheres of the Sun. Figure 7(a) presents the overall correlation between the green-line emissions in the northern and southern hemispheres, recording a 0.93 correlation coefficient without recording a lag or lead. The significant peak at zero lag indicates synchronized green-line emission behaviors. However, the correlation is higher on the right wing, demonstrating that emissions in one hemisphere might lead those in the other by a small amount of time. This asymmetric distribution may be attributed to the uneven distributions of activities relating to magnetic fields or the influence of large-scale solar dynamics affecting one hemisphere slightly earlier than the other.

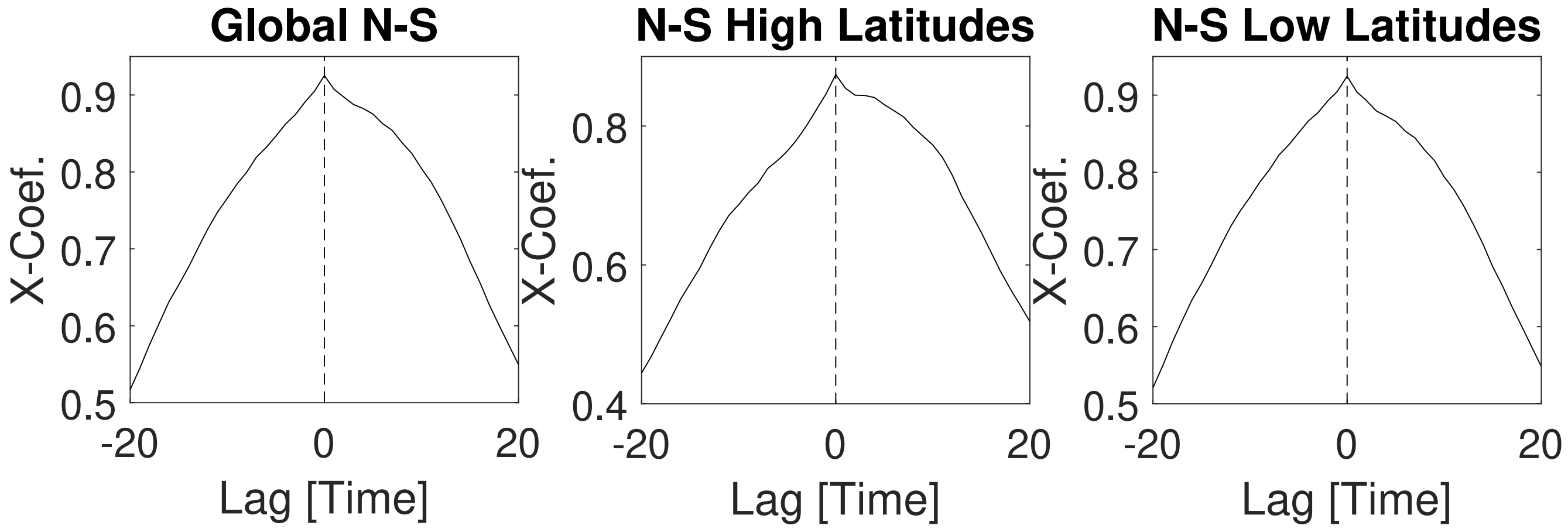

**Figure 7.** The cross correlations between the solar north and south regions.

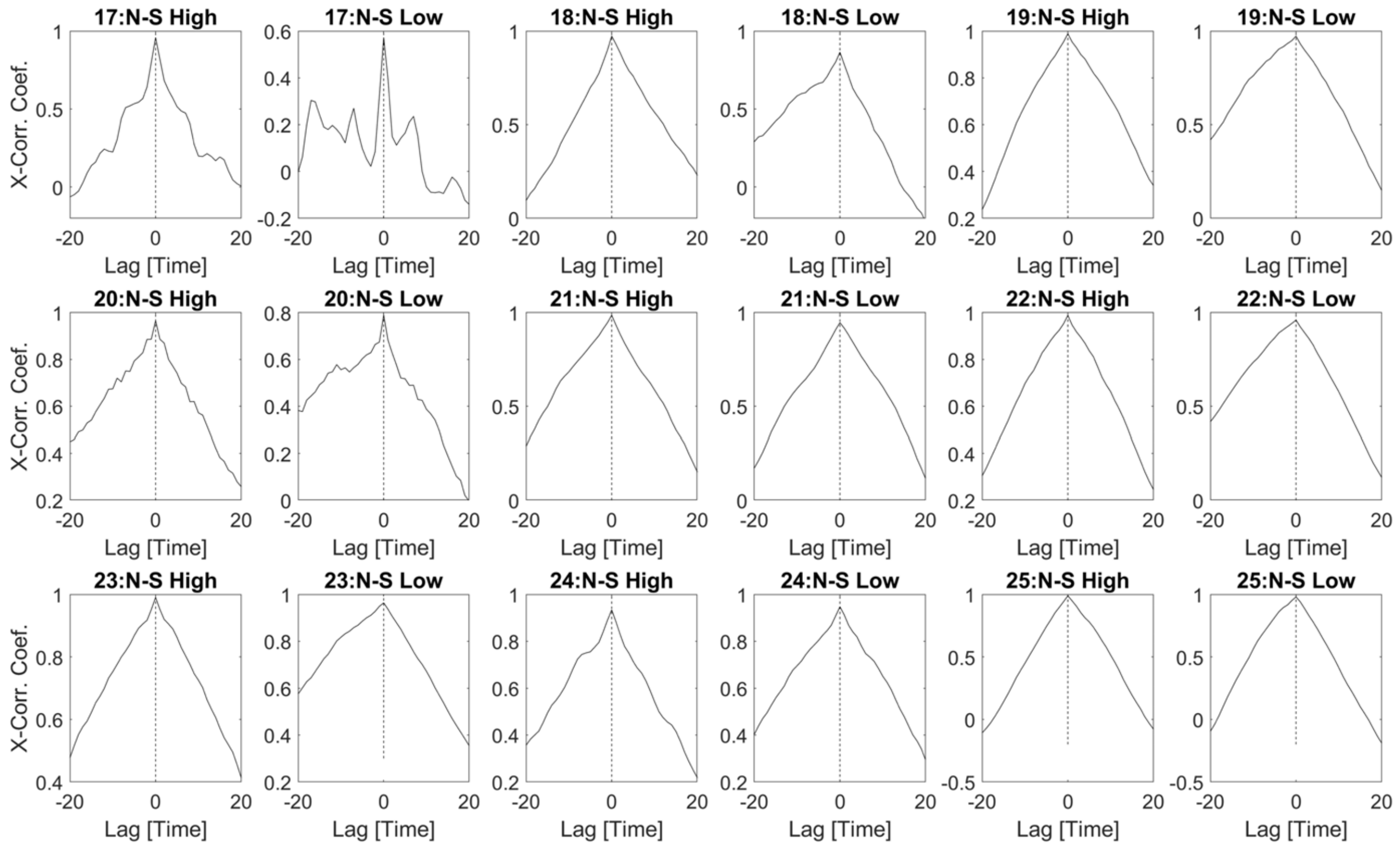

**Figure 8.** The cross correlations between the solar north and south regions across the solar cycles.

At high latitudes presented in Figure 7(b), it shows a correlation coefficient of 0.87, with a more pronounced asymmetry. The right wing again shows higher correlation values compared to the left wing. This implies that high-latitude features, such as polar crown prominences or polar coronal holes, might influence the opposite hemisphere after an inevitable delay. This could be due to the propagation of waves or magnetic reconnections occurring preferentially in one hemisphere and affecting the other later. B. A. Emery et al. (2021) illustrated how secondary polarity inversion lines and polar coronal hole boundaries show delayed and more poleward movement in the southern region compared to the northern region, presenting a possible linkage between high-latitude features in one hemisphere and their impact on the opposite hemisphere. The study revealed that the delay could be indicative of a wider solar dynamic, where magnetic or plasma structures at high latitudes in one hemisphere eventually affect the other hemisphere, potentially influencing solar activity and coronal behavior across the entire Sun. In the low latitudes, as illustrated in Figure 7(c), the cross correlation records a coefficient of 0.92, with the correlation curve also showing a higher positive lag side. The effect is most likely driven by the emergence of active regions, sunspots, and flares, which have a subsequent impact on the opposite hemisphere. The uneven distribution of sunspot activity and the differential rotation speed of the Sun could be factors in the observed pattern.

**Table 2**
The N–S Asymmetry Showing the Means and the Dominance Region

| Solar Cycle | Begins | Ends | Length (yr) | X-Corr-Coeff. | Lag | Sig. Level |
|---|---|---|---|---|---|---|
| | | | Global Scale | | | |
| Global Sun N–S | 1939.1 | 2024.3 | 85 | 0.93 | 0 | V. high |
| N–S (high latitude) | 1939.1 | 2024.3 | 85 | 0.87 | 0 | V. high |
| N–S (low latitude) | 1939.1 | 2024.3 | 85 | 0.92 | 0 | V. high |
| | | | Solar Cycle | | | |
| SC 17: N–S (high latitude) | 1939.1 | 1944.1 | 5.0 | 0.96 | 0 | V. high |
| N–S (low latitude) | 1939.1 | 1944.1 | 5.0 | 0.57 | 0 | V. high |
| SC 18: N–S (high latitude) | 1944.2 | 1954.2 | 10.2 | 0.97 | 0 | V. high |
| N–S (low latitude) | 1944.2 | 1954.2 | 10.2 | 0.86 | 0 | V. high |
| SC 19: N–S (high latitude) | 1954.3 | 1964.8 | 10.5 | 0.99 | 0 | V. high |
| N–S (low latitude) | 1954.3 | 1964.8 | 10.5 | 0.97 | 0 | V. high |
| SC 20: N–S (high latitude) | 1964.9 | 1976.4 | 11.4 | 0.97 | 0 | V. high |
| N–S (low latitude) | 1964.9 | 1976.4 | 11.4 | 0.79 | 0 | V. high |
| SC 21: N–S (high latitude) | 1976.5 | 1986.7 | 10.5 | 0.99 | 0 | V. high |
| N–S (low latitude) | 1976.5 | 1986.7 | 10.5 | 0.95 | 0 | V. high |
| SC 22: N–S (high latitude) | 1986.8 | 1996.8 | 9.9 | 0.99 | 0 | V. high |
| N–S (low latitude) | 1986.8 | 1996.8 | 9.9 | 0.96 | 0 | V. high |
| SC 23: N–S (high latitude) | 1996.9 | 2008.8 | 12.3 | 0.99 | 0 | V. high |
| N–S (low latitude) | 1996.9 | 2008.8 | 12.3 | 0.96 | 0 | V. high |
| SC 24: N–S (high latitude) | 2008.9 | 2019.0 | 11.0 | 0.93 | 0 | V. high |
| N–S (low latitude) | 2008.9 | 2019.0 | 11.0 | 0.95 | 0 | V. high |
| SC 25: N–S (high latitude) | 2019.1 | 2023.8 | 5.2 | 0.99 | 0 | V. high |
| N–S (low latitude) | 2019.1 | 2024.3 | 5.2 | 0.98 | 0 | V. high |

Figure 8 presents the cross correlations between the solar north and south regions across the solar cycles. The analysis shows that no region leads or lags in the evolution of green-line emissions. The correlation coefficients are significantly high across the cycles, indicating a smooth level of phase synchronizations. However, several peaks are observed in Solar Cycle 17, especially for the low latitudes, which may be translated to irregular phase relationships fluctuating between negative and positive. The highest correlation coefficient was recorded for high latitudes in Solar Cycles 19, 21, 22, 23, and 25 at 0.99, closely followed by 0.98 in the current Solar Cycle 25 (at low latitudes), and then 0.97 in Solar Cycles 18 (high latitudes), 19 (low latitudes), and 20 (high latitudes). The lowest coefficients observed were in Solar Cycle 17 (low latitudes) at 0.57, followed by 0.79 recorded in Solar Cycle 20 at low latitudes. In general observation, the level of relationship and synchronization of green-line emissions from both regions are significantly high.

The correlation coefficients obtained peak at zero lags, indicating strong synchronous relationships between the emissions from both hemispheres. As seen from the figure, the left wing, which represents the negative corrections, is slightly higher than the right wing for low latitudes across all the cycles. The difference is significant for Solar Cycles 18, 19, 20, and 23. This is similar for the high latitudes in Solar Cycles 20–24, while the right wing is observed to be slightly higher in high latitudes in Solar Cycles 17, 18, 19, and 25. This occurs in most cycles dominated by northern emissions, as observed in Table 1 and Figure 6. The right wing, which represents the positive correction, differs slightly from the left with higher coefficients, suggesting asymmetry in the timing or intensity of events between hemispheres.

The asymmetries, particularly the stronger right wings, suggest a tendency for the northern hemisphere to lead or influence the southern hemisphere more than vice versa. High-latitude correlations have often shown stronger and more asymmetric relationships compared to low-latitude correlations. The degree of asymmetry and correlation strength varies across solar cycles, indicating that the relationship between hemispheres evolves over time. The difference observed between the right and left wings suggests a lag in energy transfer and magnetic reconfiguration between the hemispheres. This indicates that processes such as magnetic reconnection, which can occur in one hemisphere, might propagate and influence the opposite hemisphere with a delay. The asymmetric behavior has also reflected variations in the solar cycle phases as observed in Table 1 and Figures 6(c) and (d). During different phases of the solar cycle, classified as rising and declining phases, the emissions show variations that could be attributed to the emergence of solar active regions and the distribution of coronal holes, which may lead to a time-dependent correlation between the hemispheres. B. A. Emery et al. (2021) revealed that latitude variations in coronal hole boundaries and polarity inversion lines are linked to solar magnetic field variations, such that the secondary polarity inversion lines typically reach higher latitudes later and more poleward in the southern hemisphere compared to the northern hemisphere.

This asymmetry is noteworthy, as it indicates that the evolutionary trends and advancement of solar phenomena, such as flares and green-line emissions, do not occur simultaneously in both hemispheres. Instead, there is a delay or advancement in one hemisphere compared to the other. This discovery holds significant importance in comprehending the solar corona. The emissions of the green-line corona are impacted by the magnetic field and plasma density, which are, in turn, affected by solar activity such as flares. The observed uneven distributions may offer valuable insights into the mechanics of energy transfer, magnetic field reconfigurations, and plasma heating processes occurring within the corona. Accurate data are essential for forecasting solar weather patterns and comprehending the fundamental mechanisms that control solar cycle dynamics.

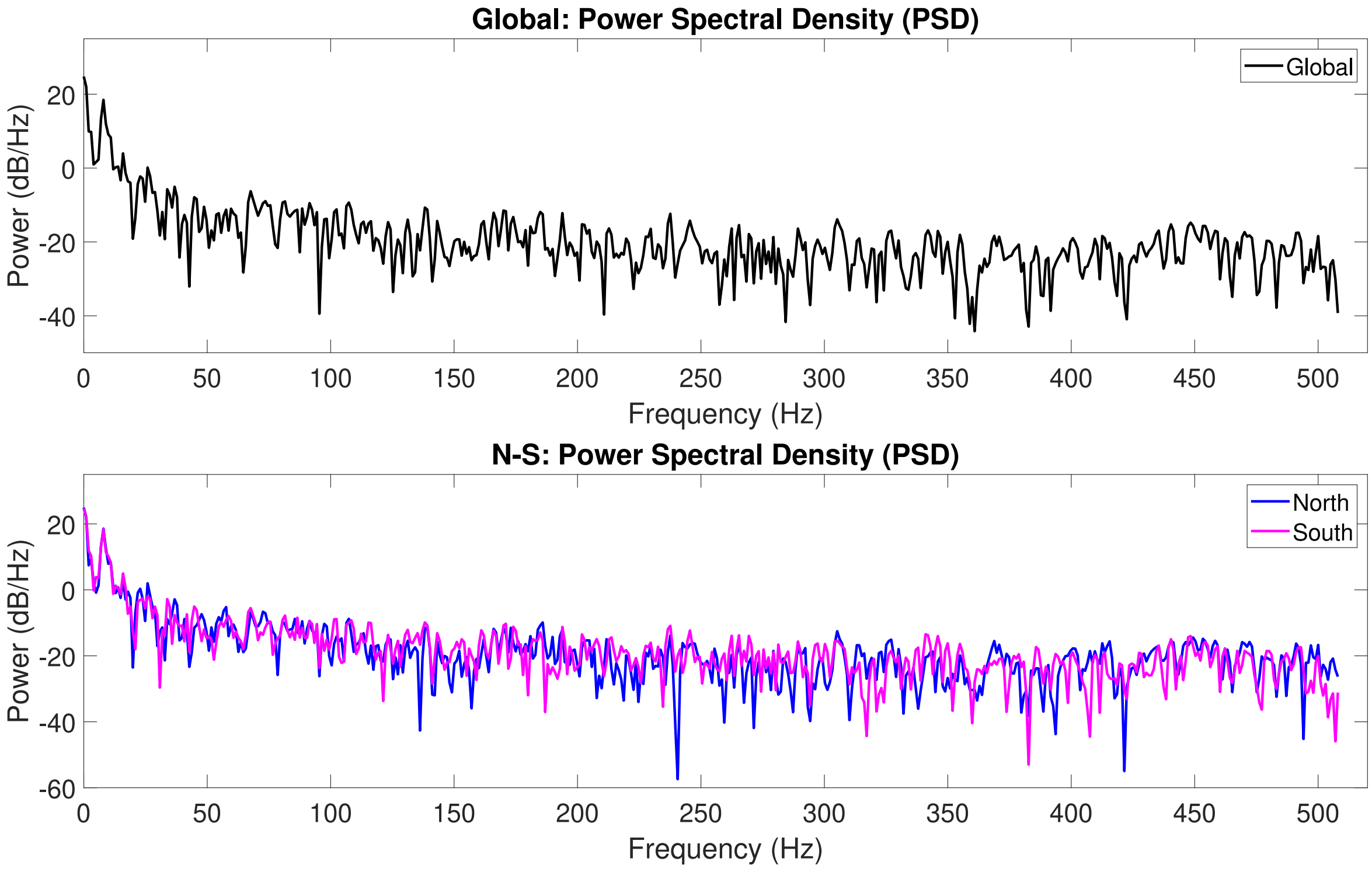

**Figure 9.** The PSD analysis of green-line coronal emissions for the global Sun (top) and the solar north (blue) and south (magenta) (bottom).

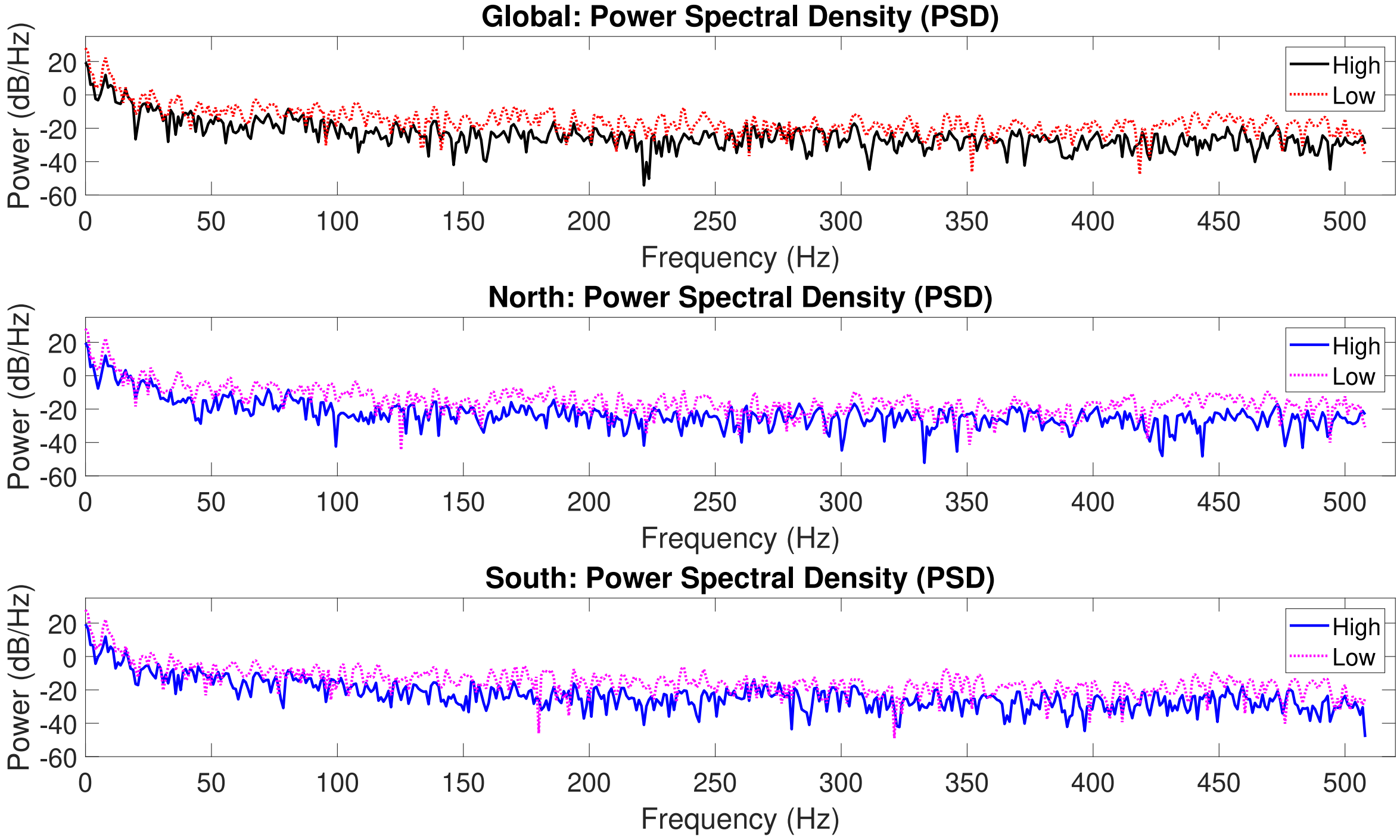

**Figure 10.** The PSD analysis of green-line coronal emissions at the high and low latitudes for (the global Sun (top), solar north (middle), and solar south (bottom).

### *3.5. The Power Spectral Density*

This method was employed because the spatial and temporal characteristics of the coronal green-line emission can be effectively analyzed using the PSD, which offers significant insights into the temporal fluctuations and periodic patterns observed in the emission data. The presence of a negative sign on the power *y*-axis does not indicate that the power is inherently negative. It depicts power levels that are lower than the reference value on the logarithmic scale. Figures 9–11 present the analyses obtained from the PSD. Figure 9 presents the PSD analysis of green-line coronal emissions for the global Sun in the top panel and the solar north (blue) and south (magenta) in the bottom panel. As seen in the figure, the distribution of power or energy of the green-line coronal emissions throughout the frequency spectrum. The emission data provide insights into the dynamics of the solar corona and associated processes that are related to its temporal variations and periodic patterns. The north–south analysis in the bottom panel of Figure 9 shows nearly the same energy/power. The energy in the north shows several significant sharp drops observed at 136, 157, 240, 259, 271, 287, 294, 310, 332, 393, 421, and 494 Hz, while it is observed in the southern region at 30, 121, 180, 186, 234, 317, 321–324, 351, 351, 359, 382, 407, 411, 438, 460, 477, and 507 Hz. These energy fluctuations correlate to the power densities that are associated with the spectral features at different frequencies.

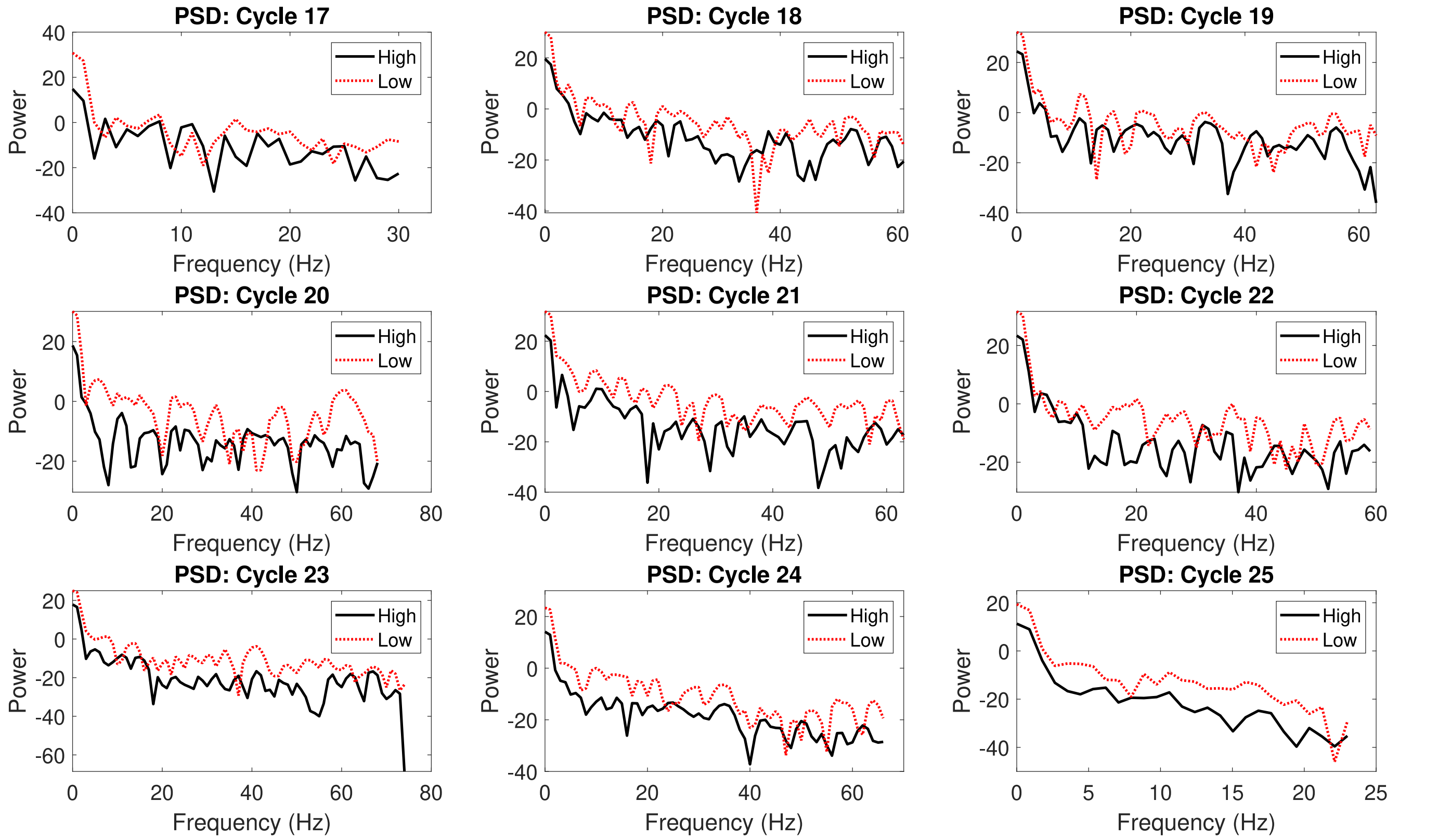

**Figure 11.** The PSD analysis of green-line coronal emissions between the solar north and south regions across the solar cycles.

The PSD analyses for the high and low latitudes are presented in Figure 10. The top panel presents the global Sun's high and low latitudes, represented by black and red colors, respectively. The middle and bottom panels show the emissions at high and low latitudes for the north and south regions of the Sun, respectively. The blue and pink colors correspond to high and low latitudes, respectively. As seen in the figure, the energy at the low latitudes remains higher at nearly all frequency spectra in the examinations conducted across the global, north, and south regions. Figure 11 presents the PSD analysis for all solar cycles. The power observed in cycle frequency rates is quite irregular, with peaks across the cycles. However, the low-latitude emissions consistently show greater power density with fewer significant drops, while high-latitude emissions are characterized by lower power density with more significant drops. Each solar cycle exhibits unique spectral characteristics, highlighting the dynamic nature of green-line emissions in the solar atmosphere.

The energy at low latitudes predominates over that at high latitudes. Notably, convergence of emission energies is observed during multiple solar cycles across various frequencies, with the exception of Solar Cycle 25, where convergence occurred twice at 8 and 22 Hz. Given the variability in the solar magnetic field and differential rotation, this energy convergence could be attributed to energy measurements taken near specific and close latitudinal boundaries, such as 220° in the high latitudes and 230° in the low latitudes.

In analyzing PSD, especially in machines, sharp drops can occur for various physical or instrumental reasons, such as destructive interference, natural resonances, or filtering effects that attenuate specific frequencies. In the context of green-line corona emissions, the sharp drops observed could be manifestations of some key factors or indicators influenced by coronal structures and magnetic field interactions at certain frequencies. Key indicators of physical processes in the corona include resonant absorption, destructive interference, filtering by coronal structures, or magnetic field dynamics. M. Luna et al. (2019) described the collective oscillations that occur during complicated interactions, where not all strands participate uniformly. This uneven participation can result in destructive interference among these oscillations, causing some frequencies to be dampened. This phenomenon is seen in the PSD analysis. Furthermore, the study also revealed that changes in the magnetic field configuration could lead to the suppression of certain frequencies. The coupling between the strands is influenced by the surrounding magnetic field, which can dynamically alter the vibrational modes. These sharp drops could also be linked to the rapid damping of the oscillations, as observed by I. De Moortel et al. (2002), which may be influenced by coronal structures and magnetic field interactions, such as sharp drops in the PSD of the green-line emissions at certain frequencies, observed in Solar Cycles 18 and 25, indicating that destructive interference is occurring in the coronal loops, possibly revealing details about the alignment and interaction of magnetic structures in the corona.

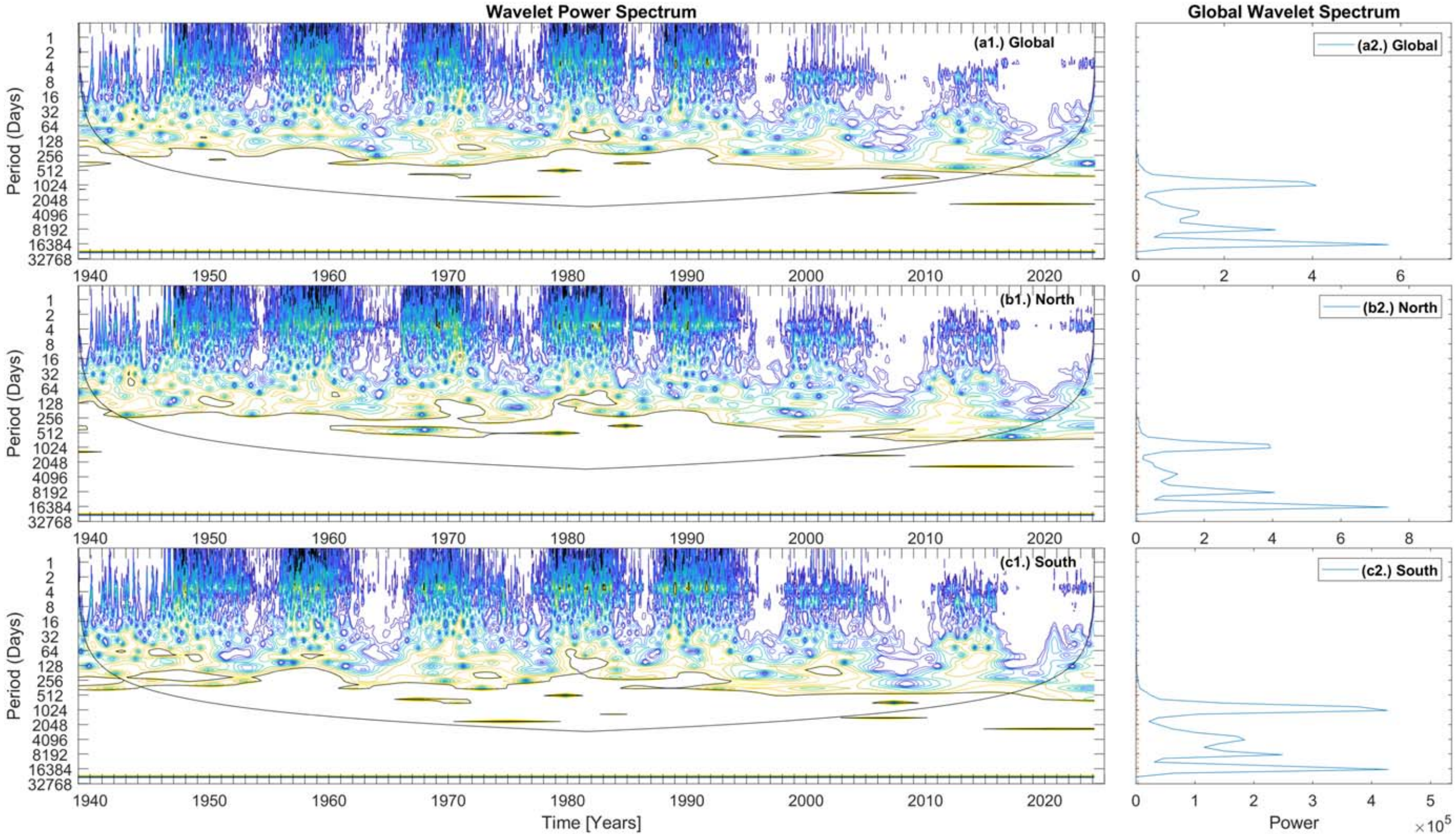

Figure 12. The wavelet analysis of green-line coronal emissions for (a) the global Sun, (b) the solar north, and (c) the solar south.

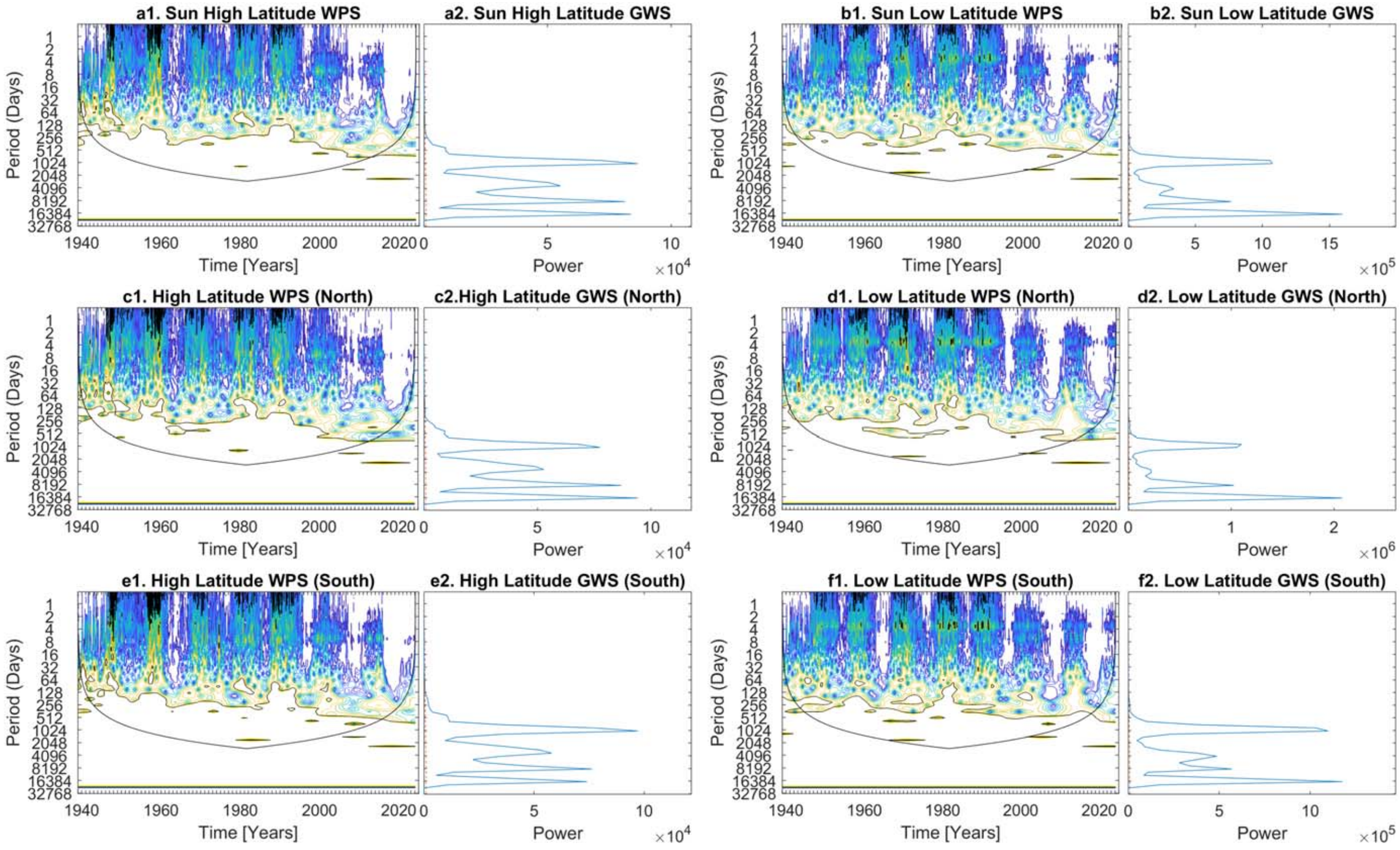

Figure 13. The wavelet analysis of green-line coronal emissions for high and low latitudes of the Sun.

The PSD analysis accurately represents changes in green-line emission over different timescales and latitudinal distributions, as it examines the distribution of power across various frequencies to identify both short-term fluctuations and long-term trends.

#### 3.5.1. The Continuous Wavelet Transform

The scales and nature of periodic variations present in the green-line coronal emissions were investigated using the Morlet CWT. The data set was subjected to the red-noise approximation. Figures 12 and 13 display the findings. The WPS and global wavelet spectrum (GWS) are also presented for each analysis. The color representations depicted in the figures provide insights into the power and frequency levels within the WPS. Low power is shown by the blue areas, while regions of higher power are indicated by the yellow and black portions. The cone of influence (COI) in each of the panels is shown by a thin black line. The purpose of introducing the COI is to reduce errors in the edges/regions that are prone to effects caused by discontinuity. Then, it is padded with zeros. The analyses are carried out at a 95% confidence level. The GWS illustrates the fluctuation of the power over the time of the investigation. The periodicity and other details obtained from the WPS and GWS are presented in Tables 3–6. The wavelet analysis for the global Sun is seen in Figure 12. The periodicity observed is located mostly between Solar Cycles 18 and 22. The observed periodicities for the global Sun are listed in Table 3, and they include 5 yr, 2–2.2 yr, 1.7–1.8 yr, 1.4 yr, and 0.98–1.26 yr. Other periodicities observed include 317–338 days, 154–205 days, 115 days, and 77–128 days.

The GWS, which presents the global view of the overriding modes of variability of the green-line data analyzed, shows periodicities of 44.89, 22.44, 11.22, and 2.81 yr at various degrees of powers, with 44.89 yr recording the most significant dominant power. This is presented in Figure 12(a). We present details of this analysis in Table 4. The analyses for the north and south regions are presented in panels (b) and (c) of the same figure. We observed periodicities of 1.68–1.82 yr, 1.4 yr, 1.12–1.68 yr, 1.0–1.12 yr, 102–154 days, 80–102 days, 45–58 days, and 14–15 days for the northern hemisphere, while the southern region shows periodicities of 5 yr, 3.7 yr, 2.0–2.24 yr, 1.68–2.24 yr, 1.4 yr, 180–230 days, 167–256 days, 137–205 days, 85–110 days, 79–99 days, 64–204 days, 64 days, and 60–85 days at various degrees of dominant powers. The details are provided in Table 3.

To gain further insights into the behaviors, distributions, and spatiotemporal periodicity of the green-line emission, we examined the process across the high- and low-latitudinal distributions across the northern and southern domains. The WPS analyses show that periodicity around 5, 2, 1.4, 1, and 0.98–1.12 yr is prominent. Other significant variations observed include 60 days, 64–128 days, 256 days, and others. We provide the details of our analyses in Figure 13, while the GWS is presented in Table 5. The analysis points to the presence of periodicity on different timescales ranging from days to years, reflecting the complex and dynamic behavior of the green-line emissions in the Sun's corona. For instance, the periodicity of 5 yr is well represented in low and high latitudes across the northern and southern hemispheres, while periods of 0.98–1.12 yr and 64–128 days are well represented in low latitudes across both regions. The GWS shows the dominant power of 44.89 yr as the highest across all the latitudes and regions. The GWS shows the harmonic behavior of green-line emissions with periodicities of 44.89, 22.44, 11.22, and 2.81 yr. This feature is also well distributed across latitudinal and regional parts of the Sun. The periodicity of 40–45 yr is not entirely reliable, as it is based on a data series spanning 85 yr. F. R. Zhu & H. Y. Jia (2018) observed 11 yr periodicity and 5.5 yr periodicity using Lomb–Scargle and New International Sunspot Numbers. A periodicity of around 44 yr has been observed, e.g., Y. Singh & Badruddin (2014) found 38.6 yr in sunspot numbers. A. Vecchio & V. Carbone (2009) observed periodicity of 1.5–4 yr early on.

The findings from O. G. Badalyan & V. N. Obridko (2006) align with our observations that green-line coronal intensity varies with latitude and solar cycle phase. While the negative correlation at high latitudes during solar minimum suggests that large-scale magnetic fields, which predominate in these regions, may suppress green-line emission, the positive correlation in the equatorial zone during solar minimum suggests that intense magnetic fields lead to increased coronal brightness. M. J. Aschwanden (2004) analyzed the temporal and spatial (equatorial and polar areas) variations in coronal heating and related phenomena. The study revealed that distinct heating processes prevail at various latitudes and throughout different stages of the solar cycle. Magnetic reconnection is likely to be more prominent during periods of solar maximum when more active regions are present. In contrast, wave dissipation may have a more significant influence during solar minimum. Previous studies conducted by J. V. Hollweg (1985) illustrated that variations in the intensity of the corona, particularly at various latitudes, can be related to differences in the efficiency of viscous heating. During moments of heightened solar activity, characterized by a more intricate and chaotic magnetic field structure, the influence of viscosity may become more prominent, resulting in elevated temperatures in the corona. On the other hand, the study shows that during solar minimum the role of viscosity in heating the corona may decrease, which is associated with lower intensities of the green line. L. Ofman et al. (1995) emphasized the crucial function of Kelvin–Helmholtz instability in intensifying turbulence and dissipation in the corona, resulting in more effective heating, particularly in areas with significant magnetic field gradients. The study demonstrates that the phase mixing of Alfvén waves in nonuniform plasmas can cause the accumulation of significant gradients' buildup, leading to concentrated heating. The interaction between shear and fast Alfvén waves in these environments intensifies the heating effect. Using planar magnetic plasma configurations, D. L. Ruderman (1997) investigated resonant shear Alfvén waves in coronal arcades. The study found solutions to viscous magnetohydrodynamic (MHD) equations, derived dissipated wave energy expressions, and computed their distribution throughout the resonant magnetic surface. G. Halberstadt & J. Goedbloed (1995) analyzed the role of the fast-wave component in Alfvén heating, focusing on the excitation of coronal loops at the footpoints due to photospheric motion. The study numerically solves the fully resistive linearized MHD equations for the model, revealing a new type of eigenmodes consisting of a global fast-wave contribution and a localized and damped Alfvén tail. In a recent investigation, Z. Lu et al. (2024) used a three-dimensional radiative MHD simulation to demonstrate how continuous magnetic flux emergence in active regions drives persistent magnetic reconnections, leading to the

**Table 3**
The Continuous Wavelet Analysis Showing the Periodicities Exhibited by the Green-line Emission in Global, North, and South Regions of the Sun

| CWT (days) | Time | Parameters | Previous Studies |
|---|---|---|---|
| | | Global Sun | |
| 5 yr | 1970–1979 | quasi-quinquennial cycle sunspots, | A. Vecchio & V. Carbone (2009), F. R. Zhu & H. Y. Jia (2018), M. Wan et al. (2020) |
| | | B-flares, solar flares | J. Oloketuyi et al. (2019), V. M. Velasco Herrera et al. (2022) |
| 2–2.2 yr | 2006–2011 | Solar flares, sunspots, solar radiation | Y. Singh & Badruddin (2014), M. Wan et al. (2020), V. V. Zharkova et al. (2023) |
| 1.7–1.8 yr | 1966–1971 | Solar wind, solar flares, sunspots | M. Wan et al. (2020), V. M. Velasco Herrera et al. (2022) |
| 1.4 yr | 1977–1981 | IMF, solar flares, solar wind | Y.-M. Wang & N. R. Sheeley, Jr (2003) |
| 0.98–1.26 yr | 1983–1987 | CME, IMF, solar wind, solar flares, sunspots | Y.-Q. Lou et al. (2003) |
| 317–338 | 1950–1951 | Solar flares, CME | Y.-Q. Lou et al. (2003) |
| 154–205 | 1970–1973 | CME, IMF, solar flares, 10.7 cm radio flux, sunspot number | E. Rieger et al. (1984), T. Bai & P. A. Sturrock (1991) |
| 115 | 1949 | Solar flares, sunspot areas, sunspot numbers, flare index | B. Joshi et al. (2006), P. R. Singh et al. (2019b) |
| 77–128 | 1980–1983 | Solar flares, sunspots | T. Bai & P. A. Sturrock (1991), T. Bai (1992, 2003) |
| | | Northern Hemisphere | |
| 1.68–1.82 yr | 1997–2001 | Solar flares | V. M. Velasco Herrera et al. (2022) |
| 1.4 yr | 1974-1981 | IMF, solar wind, geomagnetic activity index Ap, auroral index | Y.-M. Wang & N. R. Sheeley, Jr (2003), S. J. Bolton (1990) |
| 1.12–1.68 yr | 1960–1974 | CME, cosmic-ray intensity, solar wind velocity, geomagnetic activity | S. J. Bolton (1990) |
| 1.0–1.12 yr | 1983–1986 | IMF, CME, sunspots | Y.-Q. Lou et al. (2003), S. P. Nayar (2006) |
| 102–154 | 1949–1952 | Rieger cycle, IMF, solar flares, 10.7 cm radio flux, sunspot number | T. Bai & P. A. Sturrock (1991) |
| 80–102 | 1960 | Solar flares, sunspot number | T. Bai & P. A. Sturrock (1991), Y. Lou (2000), T. Bai (2003) |
| 45–58 | 1943 | F10.7 emissions, magnetic flux, sunspot | P. Chowdhury et al. (2015) |
| 14–15 | 1970–1971 | Magnetic flux, sunspot | P. Chowdhury et al. (2015), P. R. Singh et al. (2019b) |
| | | Southern Hemisphere | |
| 5 yr | 1970–1979 | Green-line emissions, sunspots, | A. Vecchio & V. Carbone (2009), J. Oloketuyi et al. (2023) |
| | | B-flares, solar flares | Y.-M. Wang & N. R. Sheeley, Jr (2003), M. Wan et al. (2020), J. Oloketuyi et al. (2023) |
| 3.7 yr | 1982-1985 | Solar flares | V. M. Velasco Herrera et al. (2022), V. V. Zharkova & S. J. Shepherd (2022) |
| 2.0–2.24 yr | 2005-2009 | CME, solar radiation, sunspots | V. V. Zharkova et al. (2023) |
| 1.68–2.24 yr | 1965-1971 | Solar flares | V. M. Velasco Herrera et al. (2022) |
| 1.4 yr | 1978–1981 | IMF, solar wind | J. D. Richardson et al. (1994), K. I. Paularena et al. (1995), Y.-M. Wang & N. R. Sheeley, Jr (2003), A. Vecchio & V. Carbone (2009) |
| 180–230 | 1972–1974 | IMF, sunspot areas | S. Sello (2003), R. Knaack et al. (2005), P. Chowdhury & B. N. Dwivedi (2011) |
| 167–256 | 1943–1952 | Sunspot area, CME | P. R. Singh et al. (2019b) |
| 137–205 | 1999–2003 | IMF, solar flares, 10.7 cm radio flux, sunspot number | Y.-Q. Lou et al. (2003), T. Barlyaeva et al. (2018) |
| 85–110 | 1948–1949 | Solar flares, sunspot number | T. Bai (2003) |
| 79–99 | 1957–1959 | Solar flares, sunspot number | T. Bai & P. A. Sturrock (1991) |
| 64–204 | 1979–1983 | CME, solar flares, solar wind | R. Knaack et al. (2005), P. Chowdhury & B. N. Dwivedi (2011), J. Oloketuyi et al. (2020) |
| 64 | 1942 | Solar flares, sunspot areas | B. Joshi et al. (2006) |
| 60–85 | 1944–1947 | Magnetic flux, sunspot, solar flares | P. Chowdhury et al. (2015), J. Oloketuyi et al. (2023) |

**Table 4**
The Global Wavelet Spectrum (in Periodic Years) Showing the Periodicities and Their Corresponding Significant Powers Analyzed in Figure 13

| Cycle Type | SG-HL GWS | $P$ ($10^4$) | SG-LL GWS | $P$ ($10^5$) | N-HL GWS | $P$ ($10^4$) | N-LL GWS | $P$ ($10^4$) | S-HL GWS | $P$ ($10^5$) | S-LL GWS | $P$ ($10^4$) |
|---|---|---|---|---|---|---|---|---|---|---|---|---|
| Hemispheric cycle (40–45 yr) | 44.89 | 8.38 | 44.89 | 16.0 | 44.89 | 9.43 | 44.89 | 20.8 | 44.89 | 7.41 | 44.89 | 11.77 |
| Hale cycle (20–22 yr) | 22.44 | 8.13 | 22.44 | 7.66 | 22.44 | 8.70 | 22.44 | 10.3 | 22.44 | 7.63 | 22.44 | 5.69 |
| Schwabe cycle (10–12 yr) | 10.10 | 5.53 | 11.22 | 3.40 | 10.10 | 5.28 | 10.10 | 2.18 | 11.22 | 5.80 | 11.22 | 4.87 |
| Quasi-biennial oscillation (QBO) (2.0–5 yr) | 2.81 | 8.66 | 2.81 | 10.74 | 2.81 | 7.75 | 2.81 | 11.0 | 2.81 | 9.72 | 2.81 | 11.00 |

**Note.** "SG," "N," "S," "LL," "HL," and "$P$" refer to Sun global, north, south, low latitude, high latitude, and power, respectively, for brevity.

**Table 5**
The Global Wavelet Spectrum Showing the Periodicities and Their Corresponding Significant Powers Analyzed in Figure 12

| Cycle Type | Global Sun | | Northern Hemisphere | | Southern Hemisphere | |
|---|---|---|---|---|---|---|
| | GWS (yr) | Power ($\times 10^5$) | GWS (yr) | Power ($\times 10^5$) | GWS (yr) | Power ($\times 10^5$) |
| Hemispheric cycle (40–45 yr) | 44.89 | 5.73 | 44.89 | 7.40 | 44.89 | 4.29 |
| Hale cycle (20–22 yr) | 22.44 | 3.17 | 22.44 | 4.06 | 22.44 | 2.49 |
| Schwabe cycle (10–12 yr) | 11.22 | 1.44 | 10.10 | 1.22 | 11.22 | 1.84 |
| QBOs (2.0–5 yr) | 2.81 | 4.09 | 2.5–2.81 | 3.94 | 2.81 | 4.27 |

formation and maintenance of superhot coronal loops. These loops are composed of numerous substructures heated independently by impulsive energy releases but collectively maintain a stable appearance over time. The model successfully explained the origin of the superhot coronal plasma and the long-lasting nature of coronal loops observed in extreme-ultraviolet and soft X-ray images.

## 4. Conclusions

In this work we have analyzed the behavior, spread, and spatiotemporal periodicity of the green-line emission using cross correlation, PSD, and wavelet transformation from 1939 to 2024 across the high- and low-latitudinal regions of the northern and southern Sun. The objective was to understand the different influences on the spatiotemporal variations and energy distributions of green-line emissions in the solar corona. According to S. P. Plunkett et al. (1997), transients in the lower corona linked to green-line emissions are related to CMEs observable in white light in the exterior corona. These CMEs are characterized by a gradually expanding loop in green-line emission, serving as the CME's leading edge. E. P. Kontar et al. (2023) offer key insights into how density turbulence affects radio burst propagation in the solar corona and heliosphere, stating that the density fluctuations decrease with heliocentric distance, demonstrating a specific trend in the solar corona. S. K. Prasad et al. (2013) noted that the width of the red emission line [Fe X] λ6374 broadened with height, while the green emission line [Fe XIV] λ5303 narrowed with increasing height. Elevated FWHM values in the polar regions were attributed to increased nonthermal velocities, likely connected to a nonthermal source driving solar wind acceleration. K. Hori et al. (2005) observed that the cutting of the tether in the lower corona and the subsequent breakout in the upper corona, preceding the breakout, collaborated to trigger CMEs. J. Huang et al. (2023) delve into various aspects of filament dynamics and emissions in the solar atmosphere, observing different emission lines and their behaviors during filament formation and evolution. The study found that bright emissions signified consistent plasma cooling at about 0.5 MK, which peaked the extent of cooler plasma and led to slower decay rates. It also determined that stable filaments and leg structures appeared as dark features, showing that lower-temperature plasma in these regions resulted in strong absorption lines, indicating emission variations based on plasma temperature differences between latitudes. This finding closely associates the solar atmosphere with variations in green-line emissions across different latitudes.

Our main findings can be summarized as follows:

1. The low latitudes exhibit higher emission rates compared to high latitudes. It was also observed that periodicity in the low latitudes does not necessarily follow the patterns exhibited in the high latitudes, as is seen across the solar cycles. At low latitudes, early rise in green-line emissions characterized the rising phase of the solar cycle as observed in Solar Cycles 19, 21, and 22.
2. It is shown that Solar Cycles 19, 21, and 22 record the highest activity of green-line coronal emission, while the lowest activity was observed in Solar Cycles 23, 24, and 25. An early rise in green-line emission activity is also noticed at high latitudes, especially in Solar Cycles 18, 19, 22, and 23. The emission activity is unusually low from the solar peak time to the final phase in Solar Cycles 17, 18, 20, and 21. This observation is also exhibited in both the northern and southern hemispheres, as shown in Figures 2(c), 3, 4, and 5. It is also observed that the emission activity at low latitudes shadows a pattern similar to the solar cycle.
3. Peaks observed in low latitudes do not follow the same trend in high latitudes, such that the emission at high latitudes displays peaks that are dissimilar to those at low latitudes.
4. The hemispheric distributions of green-line emissions show northern dominance for four consecutive solar cycles (from 17 to 20), after which it shifted southward for another four cycles, thereby establishing a 44 yr hemispheric dominance. The pattern continues to follow the same paradigm of N–S dominance with Solar Cycle 25, marking the beginning of a new cycle shift toward the north. However, it is worth noting that 85 yr of series data is only partially adequate for a reliable conclusion.
5. The reported total leads in Solar Cycles 18, 19, 21, 22, 23, 24, and 25 are mainly impacted by green-line emissions during the decline stages.
6. The average emission activity measured in the northern hemisphere includes 9.90 (high latitudes) and 27.32 (low latitudes), making emissions in the low latitudes constitute 73% of total emissions, while the emission in the southern part has 9.50 (high latitudes) and 26.20 (low latitudes), making emissions at low latitudes account for 73%. On the overall global scale, the northern average is 18.61, while the southern average is 17.85, with a northern dominance at 0.76.
7. Cycle 19 has the highest emission for the northern hemisphere and Solar Cycle 22 has the highest for the southern hemisphere at 27.55 and 25.69, respectively.
8. The correlation coefficients obtained are significantly high, demonstrating good levels of relationships. Our analysis for the global Sun recorded a 0.93 correlation coefficient without a lag or lead, demonstrating a high level of phase synchronizations between emissions from both north and south regions. The analysis for the high latitudes shows 0.87, while the analysis at the low latitudes returns 0.92.

9. The correlation coefficients obtained are all highly significant across the cycles, indicating smooth levels of phase synchronizations. The highest coefficient obtained is 0.99 in Solar Cycle 19 (high latitudes), Solar Cycle 21 (high latitudes), Solar Cycle 22 (high latitudes), Solar Cycle 23 (high latitudes), and Solar Cycle 25 (high latitudes), while the lowest, but still significant high, was obtained in Solar Cycle 17 (low latitudes) with a value of 0.57.
10. The SPD analysis provides insights into the energy fluctuations in the high and low latitudes, with several dissimilar peaks and different patterns showing the dynamics of green-line emissions and associated processes in the solar corona. It is also observed that energy in the low-latitude region sometimes exhibits sharp recession. The frequency rates are relatively irregular across the cycles; however, the low latitudes overshadow the energy at high latitudes.
11. As seen from the figures, the energy at the low latitudes remains higher at nearly all the frequency spectra. Notably, there are convergences observed across the solar cycles that could be attributed to energy captured along close latitudes between high and low latitudes.
12. The SPD analysis also accurately represents the changes in green-line emission over different timescales and latitudinal distributions, as it showed the distributions of power across various frequencies to identify both short-term fluctuations and long-term trends, with the potential to reveal the underlying physical processes that cause these variations.
13. The periodicity observed occurs mostly between Solar Cycles 18 and 22. The observed periodicity in the global Sun includes 5 yr, 2–2.2 yr, 1.7–1.8 yr, 1.4 yr, 0.98–1.26 yr, 317–338 days, 154–205 days, 115 days, and 77–128 days.
14. The periodicity observed for the northern and southern regions includes 1.68–1.82 yr, 1.4 yr, 1.12–1.68 yr, 1.0–1.12 yr, 102–154 days, 80–102 days, 45–58 days, and 14–15 days for the northern hemisphere, while the periodicity obtained in the south includes 5 yr, 3.7 yr, 2.0–2.24 yr, 1.68–2.24 yr, 1.4 yr, 180–230 days, 167–256 days, 137–205 days, 85–110 days, 79–99 days, 64–204 days, 64 days, and 60–85 days at various degrees of dominant powers.
15. The GWS, which presents the global view of the overriding modes of variability of the green-line data analyzed, shows the harmonic behavior of green-line emissions with periodicities of 44.89, 22.44, 11.22, and 2.81 yr. This feature is well distributed across latitudinal and regional parts of the Sun.
16. Periodicity of 5 yr is well represented in the high latitudes across the northern and southern hemispheres, while periods of 0.98–1.12 yr and 64–128 days are well represented across low and high latitudes of both regions.
17. The emission behaviors are shown to be strongly associated with other solar phenomena, such as solar flares, sunspots, and CMEs during the solar cycles.
18. The emission patterns in Solar Cycle 25 are already exhibiting a decrease in solar activity.

M. Temmer et al. (2006) analyzed the N–S asymmetry behavior over six solar cycles between 1945 and 2004 and found self-contained evolution of each solar hemisphere, with significant asymmetries during the solar cycle maximum and interaction during minima. The study found no systematic pattern and emphasizes the importance of studying the northern and southern hemispheres separately for solar activity (T. Bai 1992; L. H. Deng et al. 2015).

The turbulent heating and magnetic reconnection processes described by G. Inverarity & E. Priest (1995) can be linked to the observed variations in green-line coronal intensity at different latitudes. In low-latitude regions, where magnetic fields are stronger and more twisted, the frequent reconnection events and turbulent heating are likely responsible for the higher green-line intensity. In contrast, at higher latitudes, where magnetic fields may be less complex and less twisted, the reduced frequency of reconnection events could result in lower coronal heating and, consequently, lower green-line intensity. K. Galsgaard & Å. Nordlund (1996) studied coronal heating in the solar atmosphere using numerical simulations. They found that large-scale shearing motions create tangential discontinuities, or current sheets, within the magnetic field. These sheets induce a dynamic plasma state, generating supersonic and super-Alfvénic jet flows. The heating rate is influenced by factors like boundary velocity amplitude, correlation time, Alfvén speed, and magnetic field strength. The strong magnetic fields and higher plasma densities at high latitudes make them prime sites for resonant absorption and phase mixing of Alfvén waves. The enhanced turbulence and Kelvin–Helmholtz instability in these regions likely lead to more significant heating, contributing to higher green-line intensities observed at these latitudes (L. Ofman et al. 1995). While lower latitudes may have weaker magnetic fields, large-scale magnetic loops and active regions could still support resonant absorption, albeit with different characteristics. The heating at these latitudes might be less intense but could still influence the overall green-line intensity, especially during certain solar cycle phases. The coronal heating analysis conducted by O. G. Badalyan & V. N. Obridko (2007) supports the idea that different heating mechanisms dominate in different latitudinal regions. In the equatorial regions, magnetic reconnection and related processes (DC models) appear to be more significant, while at high latitudes, wave-based heating mechanisms (AC models) are likely more important. This differentiation further helps explain why green-line intensity patterns vary across the solar surface. V. M. Nakariakov et al. (2016) provided more insights into the mechanisms of coronal heating by discussing how MHD waves, particularly Alfvén waves, contribute to solar corona heating. The green-line coronal intensity, related to highly ionized iron in the corona, is an observational marker for these heating processes.

Finally, the asymmetry between the hemispheres in solar activity is not unusual. It can be influenced by various factors such as the solar magnetic field, differential rotation, and the solar dynamo mechanism. The green-line emission is a useful link for studying and understanding solar activity. Studying the spatiotemporal behavior of green-line emission and its periodicity is anticipated to provide practical insights into the solar corona dynamics, the solar magnetic field, and its formation on the solar surface. As previously stated, all continuous cycles can be valuable in formulating forecasts about solar activity (T. Bai 1992; L. H. Deng et al. 2015).

## Acknowledgments

We express our gratitude to the anonymous reviewer(s), whose comments have significantly improved our article and contributed to a deeper understanding of our work. The authors express their gratitude to the Astronomical Institute of the Slovak Academy of Sciences for providing the modified homogeneous data set (MHDS) and to the wavelet software tools provided by C. Torrence and G. Compo, http://paos.colorado.edu/research/wavelets/. This study is supported by the National Natural Science Foundation of China (NSFC 12373063, 11533009) and the 25 cm Coronagraph Development Project. J.O. is grateful for the Sichuan Province/SWJTU Postdoctoral Fellowship grant. This work is also supported by the "Yunnan Revitalization Talent Support Program" Innovation Team Project (202405AS350012) and the Yunnan Fundamental Research Projects (grant No. 202301AV070007). The research work of A.E. was supported by King Saud University's Deanship of Scientific Research and College of Science Research Center in Saudi Arabia. A.I. acknowledges the support of the National SKA Program of China with grant No. 2022SKA0110100 and the Alliance of International Science Organizations Visiting Fellowship on Mega-Science Facilities for Early-Career Scientists (grant Nos. ANSO-VF-2022-01 and ANSO-VF-2024-01).

## Appendix
## Appendix Information

Table 6 provides further evidence that emission behaviors are closely associated with other solar phenomena, such as solar flares, sunspots, and CMEs, throughout the solar cycles studied.

**Table 6**
The Continuous Wavelet Analysis Showing the Periodicities Exhibited by the Green-line Emission in High and Low Latitudes across the Northern and Southern Regions of the Sun

| CWT (days) | Time | Parameters | Previous Studies |
|---|---|---|---|
| | Sun (Global) High Latitudes | | |
| 3.65 yr | 1978–1981 | Solar flares | V. M. Velasco Herrera et al. (2022), V. V. Zharkova & S. J. Shepherd (2022) |
| 1.68 yr | 1998–2002 | Flares, solar wind | Y.-M. Wang & N. R. Sheeley, Jr (2003), S. J. Bolton (1990), V. M. Velasco Herrera et al. (2022) |
| 0.98–1.12 yr | 1970–1973 | CME, IMF, sunspots | Y.-Q. Lou et al. (2003), S. P. Nayar (2006) |
| 205–230 | 1944–1947 | Sunspot areas, magnetic field | S. Sello (2003), R. Knaack et al. (2005), P. Chowdhury & B. N. Dwivedi (2011) |
| 96–102 | 1989–1990 | Solar flares | T. Bai (2003) |
| 64–128 | 1978–1983 | Solar flares, sunspot areas, corona index | B. Joshi et al. (2006), P. Chowdhury et al. (2009) |
| 64 | 1942–1943 | Sunspot, solar flares, corona index | P. Chowdhury et al. (2009), J. Oloketuyi et al. (2020) |
| 51–60 | 1945–1946 | Solar flares, sunspot | P. Chowdhury et al. (2009) |
| 32–128 | 1945–1950 | Solar wind, solar flares, sunspot | P. Chowdhury et al. (2009), J. Oloketuyi et al. (2020) |
| 32–64 | 1939–1941 | Solar rotational modulation, solar wind | P. A. Gilman (1974), J. Oloketuyi et al. (2020) |
| 22–30 | 1945–1946 | Solar wind velocity, solar rotational modulation | S. P. Nayar (2006) |
| | Sun (Global) Low Latitudes | | |
| 5 yr | 1966–1976 | Sunspots, solar flares | J. Oloketuyi et al. (2019), V. M. Velasco Herrera et al. (2022), |
| 2.0–2.24 yr | 2006–2010 | Solar flares, sunspots, solar radiation | M. Wan et al. (2020), V. V. Zharkova et al. (2023) |
| 1.68–1.82yr | 1965–1974 | Solar flares, solar wind, green-line emissions | V. M. Velasco Herrera et al. (2022) |
| 1.4 yr | 1976–1980 | IMF, solar wind, solar flares | Y.-M. Wang & N. R. Sheeley, Jr (2003), S. J. Bolton (1990) |
| 1.05–1.12 yr | 1983–1988 | CME, solar flares, sunspots | Y.-Q. Lou et al. (2003), S. P. Nayar (2006) |
| 0.89–1.15 yr | 1959–1963 | CME, IMF, solar flares, sunspots | K. Mursula & B. Zieger (2000), A. Lopez-Comazzi & J. J. Blanco (2022) |
| 0.86–1.0 yr | 1949–1951 | CME, IMF, solar flares, sunspots | K. Mursula & B. Zieger (2000), A. Lopez-Comazzi & J. J. Blanco (2022) |
| 128–256 | 1968–1974 | Cosmic ray, CME | J. Oloketuyi et al. (2020) |
| 77–90 | 1967–1968 | Sunspot, solar flares | T. Bai (1992), Z.-Q. Yin et al. (2007) |
| 64–128 | 1980–1983 | Solar flares, sunspot, cosmic ray | B. Joshi et al. (2006), P. Chowdhury et al. (2009), J. Oloketuyi et al. (2020) |
| | North High Latitudes | | |
| 3.65 yr | 1977–1982 | Solar flares, sunspot area | V. M. Velasco Herrera et al. (2022), V. V. Zharkova & S. J. Shepherd (2022) |
| 1.82 yr | 1948–1950 | Cosmic ray, solar flares | A. Lopez-Comazzi & J. J. Blanco (2022) |
| 1.0–1.4 | 1961–1972 | IMF, solar wind, cosmic ray, solar flares, sunspots, CME | Y.-Q. Lou et al. (2003), A. Lopez-Comazzi & J. J. Blanco (2022) |
| 1.0–1.12 yr | 2012–2014 | Solar flares, sunspots, CME | Y.-Q. Lou et al. (2003), S. P. Nayar (2006) |
| 90–154 yr | 1947–1952 | IMF, Rieger cycle, 10.7 cm radio flux, sunspot number | A. Lopez-Comazzi & J. J. Blanco (2022) |

**Table 6**
(Continued)

| CWT (days) | Time | Parameters | Previous Studies |
|---|---|---|---|
| 90–115 | 1989–1991 | Sunspot, solar flares | T. Bai (1992), P. Chowdhury et al. (2009) |
| 77–180 | 1954–1960 | IMF, CME, solar flares | T. Bai & P. A. Sturrock (1991), T. Bai (1992, 2003) |
| 76–205 | 1978–1984 | Cosmic ray, solar flares, corona index | R. Knaack et al. (2005), P. Chowdhury & B. N. Dwivedi (2011) |
| 60 | 1946–1947 | Solar flares, sunspot number | P. Chowdhury & B. N. Dwivedi (2011) |
| 49–72 | 1943 | Sunspot group | P. R. Singh et al. (2019a), J. Oloketuyi et al. (2023) |
| 47–90 | 1939–1941 | Cosmic ray, solar indices | R. P. Kane (2005), P. R. Singh et al. (2019a) |
| 32–90 | 1945–1947 | Flares, solar rotational modulation (harmonics) | P. R. Singh et al. (2019a), J. Oloketuyi et al. (2019), P. Chowdhury & B. N. Dwivedi (2011) |
| 30 | 1943 | Solar rotational modulation | S. P. Nayar (2006), J. Sharma et al. (2024) |
| | | North Low Latitudes | |
| 5 yr | 1967–1975 | Flares, sunspots | F. R. Zhu & H. Y. Jia (2018), J. Oloketuyi et al. (2019) |
| 1.68–1.82 yr | 1997–1998 | Solar flares | Y.-M. Wang & N. R. Sheeley, Jr (2003), V. M. Velasco Herrera et al. (2022), |
| 1.12–1.68 | 1959–1981 | Cosmic ray, solar wind, solar flares | A. Lopez-Comazzi & J. J. Blanco (2022), V. M. Velasco Herrera et al. (2022) |
| 1.0–1.17 | 1983–1986 | IMF, solar flares, sunspots | Y.-Q. Lou et al. (2003) |
| 120–154 | 1950–1952 | Rieger cycle, IMF, solar flares, 10.7 cm radio flux, sunspot number, CME | E. Rieger et al. (1984), T. Bai (1992), J. Oloketuyi et al. (2023) |
| 83–102 | 1960–1961 | CME, cosmic ray, sunspot number and areas | Y.-Q. Lou et al. (2003), A. Lopez-Comazzi & J. J. Blanco (2022) |
| 83–90 | 1966–1968 | Solar flares | J. Oloketuyi et al. (2023) |
| 64–128 | 1981–1983 | F.10, cosmic ray, solar flares (harmonics) | R. P. Kane (2002), A. Lopez-Comazzi & J. J. Blanco (2022) |
| 13–16 | 1970–1971 | Cosmic rays, solar wind, magnetic flux | S. P. Nayar (2006), P. Chowdhury et al. (2015), J. Oloketuyi & O. Omole (2024) |
| | | South High Latitudes | |
| 5 yr | 1978–1987 | Sunspots, solar flares | I.-H. Cho & H.-Y. Chang (2011), J. Oloketuyi et al. (2019) |
| 2.0–2.24 yr | 2006–2011 | Solar flares, solar radiation | M. Wan et al. (2020), V. V. Zharkova et al. (2023) |
| 1.74–1.85 yr | 1996–1999 | IMF, solar flares, solar wind | K. Mursula & B. Zieger (2000) |
| 1.4 yr | 1980–1981 | IMF, solar wind, solar flares | Y.-M. Wang & N. R. Sheeley, Jr (2003), A. Lopez-Comazzi & J. J. Blanco (2022) |
| 1.0–1.12 yr | 1971–1974 | IMF, solar flares, sunspots | Y.-Q. Lou et al. (2003), S. P. Nayar (2006), A. Lopez-Comazzi & J. J. Blanco (2022) |
| 1.0–1.1 yr | 1962–1963 | IMF, solar flares, sunspots | Y.-Q. Lou et al. (2003), S. P. Nayar (2006), A. Lopez-Comazzi & J. J. Blanco (2022) |
| 0.89–1.1 yr | 1991–1993 | IMF, solar flares, sunspots | A. Lopez-Comazzi & J. J. Blanco (2022) |
| 246–282 | 1943–1945 | CME, sunspot areas | S. Sello (2003), Y.-Q. Lou et al. (2003) |
| 188–222 | 2001–2003 | Sunspot areas | P. R. Singh et al. (2019b) |
| 180–215 | 1940–1943 | Sunspot areas | P. R. Singh et al. (2019b) |
| 128–141 | 1966–1967 | SF10.7, solar flares, sunspot number | Y. Lou (2000), P. Chowdhury et al. (2010) |
| 115–180 | 1943–1945 | Cosmic ray, sunspot number, sunspot area | P. Chowdhury et al. (2010), P. R. Singh et al. (2019a) |
| 96–102 | 1971–1973, 74–75 | Sunspot number, solar flares | T. Bai (2003) |
| 83–128 | 2000–2002 | Cosmic ray, solar flares | Y. Lou (2000), A. Lopez-Comazzi & J. J. Blanco (2022) |
| 64–154 | 1978–1983 | Rieger cycle, cosmic ray, IMF, solar flares, 10.7 cm radio flux, sunspot number | E. Rieger et al. (1984), Y. Lou (2000), J. Oloketuyi et al. (2020) |
| 58–77 | 1941–1943 | Radio emissions, solar flares | T. Bai (1992), R. P. Kane (2002) |
| 45–115 | 1946-1950 | IMF, solar indices | P. Chowdhury et al. (2015) |
| 45–90 | 1956–1958 | Sunspot areas | P. Chowdhury et al. (2009) |
| 30–42 | 1943 | IMF | P. R. Singh et al. (2019b), J. Sharma et al. (2024) |
| 30–35 | 1957 | Rotation period, IMF, solar flares | P. R. Singh et al. (2019b), J. Sharma et al. (2024) |
| 22–30 | 1945–1946 | Sunspot, IMF, rotation period, solar rotational modulation | Z.-Q. Yin et al. (2007), J. Sharma et al. (2024) |
| 16–22 | 1947 | Solar wind, magnetic flux, sunspot | P. Chowdhury et al. (2015), J. Oloketuyi et al. (2020) |
| 11–15 | 1947 | UV flux, 10.7 cm | S. D. Bouwer (1992) |
| 6–8 | 1947–1948 | Solar wind | J. Oloketuyi et al. (2020), S. P. Nayar (2006) |
| | | South Low Latitudes | |
| 5 yr | 1966–1975 | Sunspots | I.-H. Cho & H.-Y. Chang (2011) |
| 2.0–2.17 yr | 2005–2008 | Solar flares, sunspots, solar radiation | V. V. Zharkova et al. (2023), M. Wan et al. (2020) |
| 1.74–1.85 yr | 1964–1973 | IMF, solar flares | K. Mursula & B. Zieger (2000) |
| 1.36–1.5 yr | 1976–1980 | Cosmic ray, IMF, solar wind | A. Lopez-Comazzi & J. J. Blanco (2022) |
| 0.98–1.12 yr | 1960–1962 | IMF, solar flares, sunspots | S. P. Nayar (2006), Y.-Q. Lou et al. (2003) |

**Table 6**
(Continued)

| CWT (days) | Time | Parameters | Previous Studies |
|---|---|---|---|
| 0.63–1.12 yr | 1939–1954 | IMF, solar flares, sunspots | A. Lopez-Comazzi & J. J. Blanco (2022), Y.-Q. Lou et al. (2003) |
| 154–251 | 1966–1975 | Rieger cycle, IMF, solar flares, 10.7 cm radio flux, sunspot number | P. R. Singh et al. (2019b), T. Bai (1992) P. Chowdhury & B. N. Dwivedi (2011), |
| 128–213 | 1999–2002 | Solar flares, sunspot | T. Bai (1992), P. Chowdhury et al. (2015) |
| 80–102 | 1948–1949 | Solar flare, sunspot areas | P. Chowdhury et al. (2009), T. Bai (2003) |
| 70–97 | 1957–1960 | Solar flares, X-ray | R. P. Kane (2002) |
| 64–68 | 1980 | Solar flares, sunspot areas | P. Chowdhury et al. (2009) |
| 62–73 | 1968–1969 | Solar flares | A. Ozguc & T. Atac (1994), A. Ozguc et al. (2021) |
| 60–85 | 1944–1946 | IMF, solar Indices, flares, sunspot | P. Chowdhury et al. (2015), T. Barlyaeva et al. (2018) |

## ORCID iDs

Jacob Oloketuyi https://orcid.org/0000-0003-2439-2910
Yu Liu https://orcid.org/0000-0002-7694-2454
Linhua Deng https://orcid.org/0000-0003-4407-8320
Abouazza Elmhamdi https://orcid.org/0000-0002-5391-4709
Fengrong Zhu https://orcid.org/0000-0001-6894-3425
Ayodeji Ibitoye https://orcid.org/0000-0002-0966-8598
Opeyemi Omole https://orcid.org/0000-0002-1633-3059
Feiyang Sha https://orcid.org/0000-0002-2995-070X
Qiang Liu https://orcid.org/0000-0002-8500-673X